\documentclass[3p,times]{elsarticle}

\usepackage{ecrc}

\volume{00}

\firstpage{1}

\journalname{Accepted by Journal of Web Semantics}

\runauth{}

\jid{procs}

\jnltitlelogo{Journal of Web Semantics}

\usepackage{amssymb}

\usepackage{lineno}

\usepackage[figuresright]{rotating}

\usepackage{graphicx}
\graphicspath{{./figs/}}

\usepackage{amsfonts}
\usepackage{amsmath}
\usepackage{amssymb}
\usepackage{color}
\usepackage{multirow}
\usepackage[caption=false,font=footnotesize]{subfig}
\usepackage{threeparttable}
\usepackage{xspace}
\usepackage{url}

\newcommand{\desc}{\mathtt{Desc}\xspace}
\newcommand{\ent}{\mathtt{ent}\xspace}
\newcommand{\prop}{\mathtt{prp}\xspace}
\newcommand{\val}{\mathtt{val}\xspace}
\newcommand{\score}{\mathtt{score}\xspace}
\newcommand{\globalfreq}{\mathtt{gf}\xspace}
\newcommand{\typefreq}{\mathtt{tf}\xspace}
\newcommand{\numtypefreq}{\mathtt{ntf}\xspace}
\newcommand{\localfreq}{\mathtt{lf}\xspace}
\newcommand{\inst}{\mathtt{Inst}\xspace}
\newcommand{\vfreq}{\mathtt{vf}\xspace}
\newcommand{\selfinfo}{\mathtt{si}\xspace}
\newcommand{\discr}{\mathtt{ds}\xspace}
\newcommand{\numsim}{\mathtt{ns}\xspace}
\newcommand{\pr}{\mathtt{pr}\xspace}
\newcommand{\nb}{\mathtt{Nbr}\xspace}
\newcommand{\mmr}{\mathtt{mmr}\xspace}
\newcommand{\qr}{\mathtt{qr}\xspace}
\newcommand{\simi}{\mathtt{sim}\xspace}

\newcommand{\rel}{\mathtt{rel}\xspace}

\newcommand{\frq}{\mathtt{FRQ}\xspace}
\newcommand{\exc}{\mathtt{EXC}\xspace}
\newcommand{\dsc}{\mathtt{DSC}\xspace}
\newcommand{\pagerank}{\mathtt{PageRank}\xspace}
\newcommand{\bl}{\mathtt{BL}\xspace}
\newcommand{\pmi}{\mathtt{PMI}\xspace}
\newcommand{\hits}{\mathtt{hits}\xspace}

\DeclareMathOperator*{\argmax}{arg\,max}
\newtheorem{example}{Example}

\begin{document}

\begin{frontmatter}



\dochead{}

\title{Entity Summarization: State of the Art and Future Challenges}


\author[nju]{Qingxia~Liu}
\ead{qxliu2013@smail.nju.edu.cn}

\author[nju]{Gong~Cheng\corref{cor}}
\ead{gcheng@nju.edu.cn}
\cortext[cor]{Corresponding author; tel: +86~(0)25~89680923; fax: +86~(0)25~89680923}

\author[sra]{Kalpa~Gunaratna\fnref{t1}}
\ead{k.gunaratna@samsung.com}
\fntext[t1]{Part of this work has been done while he was at Kno.e.sis Center, Wright State University.}

\author[nju]{Yuzhong~Qu}
\ead{yzqu@nju.edu.cn}

\address[nju]{State Key Laboratory for Novel Software Technology, Nanjing University, China}
\address[sra]{Samsung Research America, Mountain View CA, USA}

\begin{abstract}
The increasing availability of semantic data has substantially enhanced Web applications. Semantic data such as RDF data is commonly represented as entity-property-value triples. The magnitude of semantic data, in particular the large number of triples describing an entity, could overload users with excessive amounts of information. This has motivated fruitful research on automated generation of summaries for entity descriptions to satisfy users' information needs efficiently and effectively. We focus on this prominent topic of entity summarization, and our research objective is to present the first comprehensive survey of entity summarization research. Rather than separately reviewing each method, our contributions include (1)~identifying and classifying technical features of existing methods to form a high-level overview, (2)~identifying and classifying frameworks for combining multiple technical features adopted by existing methods, (3)~collecting known benchmarks for intrinsic evaluation and efforts for extrinsic evaluation, and~(4)~suggesting research directions for future work. By investigating the literature, we synthesized two hierarchies of techniques. The first hierarchy categories generic technical features into several perspectives: frequency and centrality, informativeness, and diversity and coverage. In the second hierarchy we present domain-specific and task-specific technical features, including the use of domain knowledge, context awareness, and personalization. Our review demonstrated that existing methods are mainly unsupervised and they combine multiple technical features using various frameworks: random surfer models, similarity-based grouping, MMR-like re-ranking, or combinatorial optimization. We also found a few deep learning based methods in recent research. Current evaluation results and our case study showed that the problem of entity summarization is still far from being solved. Based on the limitations of existing methods revealed in the review, we identified several future directions: the use of semantics, human factors, machine and deep learning, non-extractive methods, and interactive methods.
\end{abstract}


\end{frontmatter}


\section{Introduction}
\label{sec:introduction}

Semantic data has been used to broadly refer to structured or semi-structured data that allows machines to easily understand and manipulate the conveyed information. It facilitates data integration and enables applications to derive value from each other. Today, Web applications make information publicly accessible not only as human-readable web pages but also as machine-readable semantic data. For example, semantic data has been either embedded in HTML web pages using markup formats like RDFa~\cite{rdfa}
or Microdata~\cite{microdata}, or served directly as dump files or Linked RDF Data~\cite{ld}. Semantic data in the form of a knowledge graph helps enterprises to drive products and make them more intelligent~\cite{DBLP:journals/cacm/NoyGJNPT19}.

\begin{figure}[!t]
	\centering
	\includegraphics[width=\linewidth]{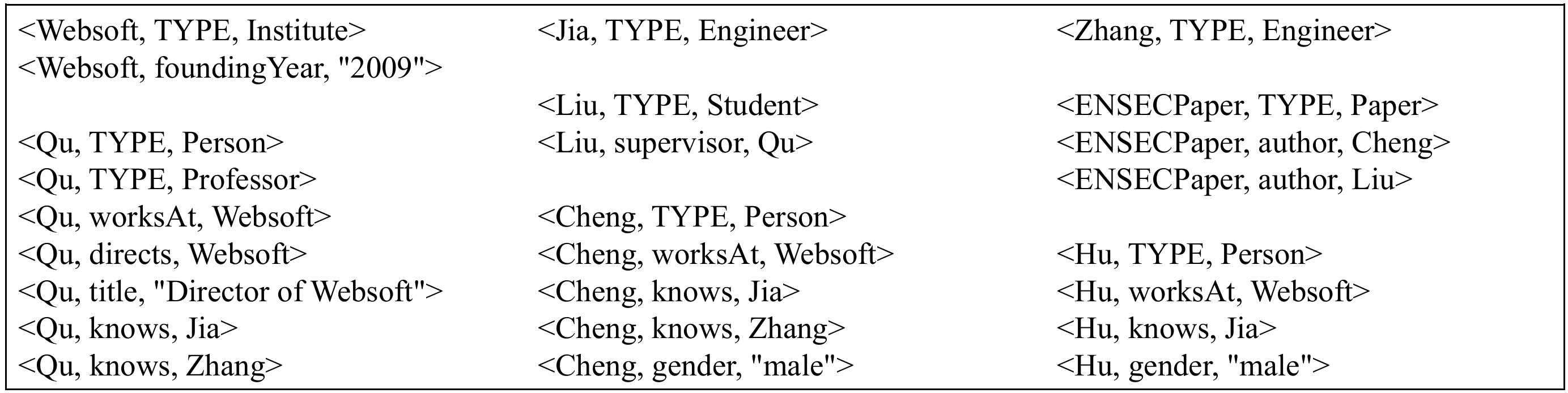}
	\caption{A running example of semantic data.}
	\label{fig:exampletable}
\end{figure}

\begin{figure}[!t]
	\centering
	\includegraphics[width=\linewidth]{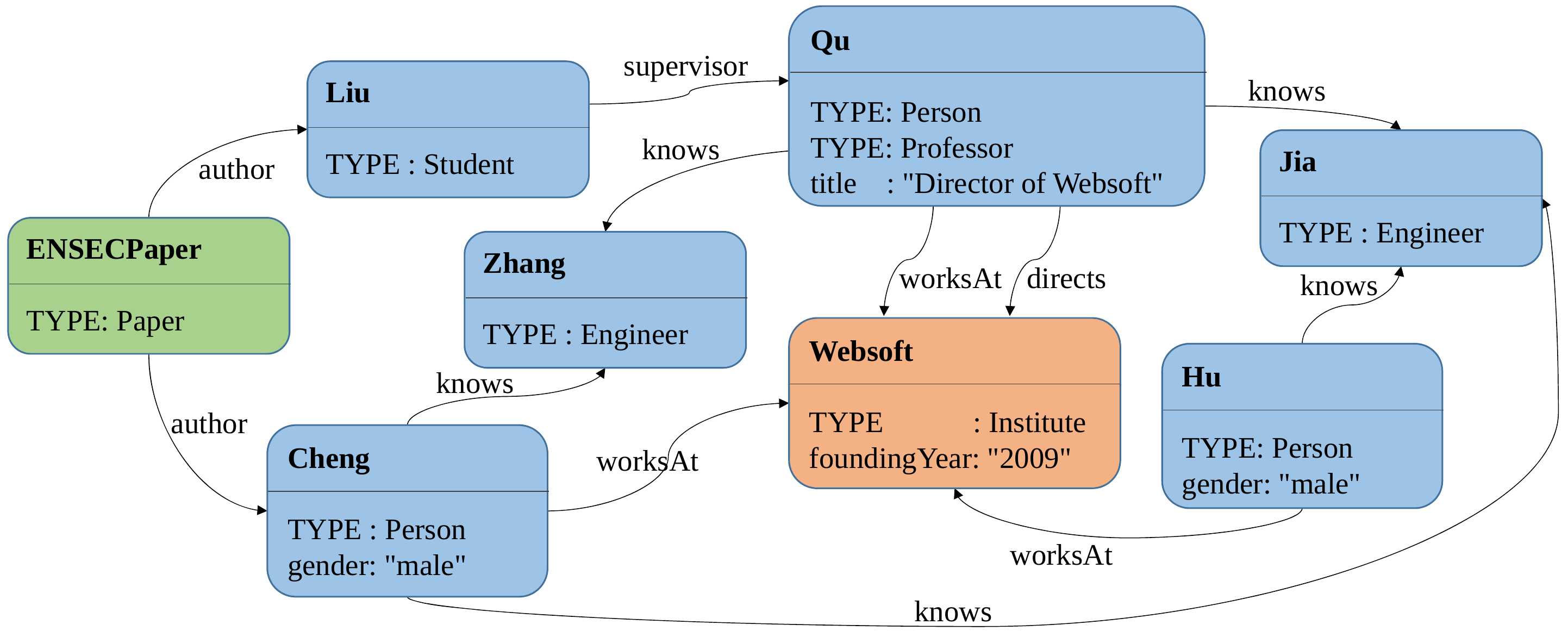}
	\caption{A graph representation of our running example of semantic data.}
	\label{fig:examplegraph}
\end{figure}

Many of these different formats of semantic data essentially adopt a common generalized data model. They describe entities with property-value pairs, which collectively form \emph{entity-property-value triples}, or \emph{triples} for short. As a running example for this article, Figure~\ref{fig:exampletable} shows a set of triples about people in a research group. An entity has one or more types such as \texttt{Person}, \texttt{Professor}, and \texttt{Engineer}. Properties are divided into three categories: \texttt{TYPE}, attributes, and relations, according to their binding values.
\begin{itemize}
	\item \texttt{TYPE} has entity types (called classes) as values.
	\item Attributes have primitive data values (called literals) as values, e.g., \texttt{title} whose value is a string.
	\item Relations have entities as values, e.g., \texttt{worksAt} whose value is an \texttt{Institute} entity.
\end{itemize}
\noindent A set of triples can be represented as a directed graph where vertices represent entities annotated with types and attributes, interconnected by arcs representing relations. For example, Figure~\ref{fig:examplegraph} depicts the triples in Figure~\ref{fig:exampletable}.

The Web has witnessed an explosive growth of semantic data over the past few years. As of September 2020, more than 86~billion triples had been embedded in 50\% of HTML web pages and found in 15~million domains\footnote{\url{http://webdatacommons.org/structureddata/}}. As of May 2020, more than one thousand datasets are available as Linked RDF Data, covering government, life sciences, publications, social networking, and other domains\footnote{\url{https://lod-cloud.net/}}. Besides, knowledge graphs comprising tens of billions of triples have been built in some of the largest technology companies~\cite{DBLP:journals/cacm/NoyGJNPT19}.

\subsection{Motivation and Application for Entity Summarization}

Semantic data, though primarily aiming to be consumed by machines, is sometimes exposed to human users in a fairly plain format. For example, to show an entity mentioned in a search query, Google Search retrieves from its Knowledge Graph a set of triples describing the entity. These triples are shown on the right-hand side of Google's search results pages~\cite{kg}. However, the description of an entity may comprise dozens or hundreds of triples, exceeding the capacity of a typical user interface---particularly on mobile devices. A user, if being served with all of those triples, would suffer information overload and find it difficult to quickly identify the small subset of facts that are truly needed.

To address this problem, one promising solution is to not show the entire entity description but \emph{provide a short summary for the entity}. A carefully generated summary, though possibly providing less and incomplete information, has the potential to cost-effectively satisfy a user's information need. For example, instead of showing all the triples describing an entity in the Knowledge Graph, Google provides ``the best summary'' for the entity by choosing and presenting a subset of triples that the user is likely to be searching for that particular entity~\cite{kg}. This problem of automatically summarizing entity descriptions has been termed \emph{entity summarization} by the research community.

In addition to search engines, entity summarization supports a multiplicity of other applications~\cite{thesis1,thesis2}. For example, entity summarization facilitates document browsing. Some applications adopt entity linking techniques~\cite{el} to enrich the content of a document (e.g.,~a news article) by linking entity mentions in text with their corresponding entities in semantic data. Triples describing these entities are then extracted from semantic data, to provide additional information to the document and allow exploratory browsing options that are relevant to the document content. Entity summarization helps to control the number of triples presented to users at a manageable level and provide the most useful triples~\cite{trank,eles,ijcai17}.

Entity summarization has also assisted many research activities. For example, in crowdsourced entity linking~\cite{zencrowd}, human participants manually link entity mentions in text to entities in semantic data. After seeing an entity mention, a participant retrieves a set of candidate entities from semantic data. Each of them is described by possibly a large set of triples. Entity summarization retains the most distinctive triples, and helps the participant quickly identify the correct entity~\cite{summel}. Analogously, by identifying the most similar triples, entity summarization has facilitated human intervention in entity resolution to quickly determine whether two entity descriptions refer to the same real-world object~\cite{DBLP:conf/esws/XuCQ14,c3dp,ctab}. In these activities, human participants complete their tasks more efficiently when they use summaries of entity descriptions.

\subsection{Problem Statement}
\label{sect:ps}

We present a generalized definition of entity summarization that is compatible with many existing research efforts. Our definition is independent of the concrete processed data format and targeted application scenario.

\paragraph{Semantic Data}

Let $\mathbb{E},\mathbb{P},\mathbb{C},\mathbb{L}$ be the sets of all entities, properties, classes, and literals, respectively. Properties are divided into $\texttt{TYPE} \in \mathbb{P}$, attributes~$\mathbb{A} \subseteq \mathbb{P}$, and relations~$\mathbb{R} \subseteq \mathbb{P}$. Semantic data is a set of entity-property-value triples denoted by $T \subseteq \mathbb{E} \times \mathbb{P} \times (\mathbb{C} \cup \mathbb{L} \cup \mathbb{E})$, or more precisely:
\begin{equation}
    T \subseteq (\mathbb{E} \times \{\texttt{TYPE}\} \times \mathbb{C}) \cup (\mathbb{E} \times \mathbb{A} \times \mathbb{L}) \cup (\mathbb{E} \times \mathbb{R} \times \mathbb{E}) \,.
\end{equation}

For readers who are familiar with Semantic Web standards, an entity-property-value triple can be represented as a subject-predicate-object triple in RDF~\cite{rdf}, which is a framework for representing information in the Web and has a graph-based data model. \texttt{TYPE}, attributes, and relations correspond to \texttt{rdf:type}, data properties, and object properties, respectively. However, we do not use Semantic Web terms in this article because entity summarization is not specific to semantic data that complies with Semantic Web standards.

Note that in RDF, property values can also be blank nodes which often represent anonymous entities/classes or n-ary relations. A blank node is not a self-contained value but relies on other triples describing it. Triples involving blank nodes are difficult to handle and hence are usually ignored by current research on entity summarization.

\paragraph{Entity Description}

For convenience, let $\ent(t), \prop(t), \val(t)$ return the entity, property, and value in a triple $t \in T$, respectively. In semantic data~$T$, the description of an entity~$e \in \mathbb{E}$ consists of the subset of triples where $e$~is described as an entity or as a property value:
\begin{equation}
    \desc(e) = \{t \in T : \ent(t)=e \text{ or } \val(t)=e\} \,.
\end{equation}
\noindent For example, in Figure~\ref{fig:exampletable}, the description of entity \texttt{Qu} comprises eight triples, including $\langle$\texttt{Liu}, \texttt{supervisor}, \texttt{Qu}$\rangle$ where \texttt{Qu} is the property value. For convenience, when the description of \texttt{Qu} is to be summarized, we re-write the above triple as $\langle$\texttt{Qu}, \texttt{supervisor}$^-$, \texttt{Liu}$\rangle$ where \texttt{supervisor}$^-$ represents the inverse of \texttt{supervisor}, and \texttt{Liu} becomes the property value. In this way, in $\desc(e)$ which is to be summarized, every triple has~$e$ at the entity position. Summarization is then focused on the property-value pair represented by each triple.

\paragraph{Entity Summary}

Existing efforts to summarize entity descriptions are mainly extractive solutions. They define a summary~$S$ of entity~$e$ as a size-constrained subset of triples selected from the description of~$e$. A size constraint is usually an upper bound~$k$ on the number of selected triples,
i.e.,~a summary $S \subseteq \desc(e)$ satisfies $|S| \leq k$. For example, given $k=5$, any 1--5~triples selected from the description of~\texttt{Qu} in Figure~\ref{fig:exampletable} form a summary of~\texttt{Qu}.

\paragraph{Entity Summarization}

The problem of entity summarization is formulated as finding an optimal summary:
\begin{equation}\label{eq:problem}
    \text{find } \argmax_{S \subseteq \desc(e)}{\score(S|T)}, \quad \text{ subject to } |S| \leq k \,,
\end{equation}
\noindent where $\score(S|T)$ is the quality score of summary~$S$ given~$T$. Note that the score of~$S$ is conditioned on~$T$. Although we focus on selecting triples from~$\desc(e)$, all the triples in~$T$---including those outside~$\desc(e)$---are considered as input because they are often useful when assessing triples in~$\desc(e)$.

An algorithm or system for solving the problem of entity summarization is called an entity summarizer. Different entity summarizers define and compute~$\score(S|T)$ in different ways. Some reduce Eq.~(\ref{eq:problem}) to a ranking problem, e.g.,~\cite{relin,usewod,waim}. They assume
\begin{equation}\label{eq:problemrank}
    \score(S|T) = \sum_{t \in S}{\score(t|T)} \,,
\end{equation}
\noindent where $\score(t|T)$ is the quality score of triple $t \in S$. So the problem turns into ranking the triples in $\desc(e)$ and then generating a summary by choosing $k$~top-ranked triples. Ranking is also a major step in many other entity summarizers, e.g.,~\cite{faces,facese,ensec2,linksum}. As entity summarization is closely related to triple ranking, we will not particularly make a distinction between summary-based methods and ranking-based methods.

\subsection{Scope of Survey}

The identification of entity summarizers for our survey was done according to the following strategy.

First, we scanned the proceedings of a set of relevant conference series and the volumes of some relevant journals starting from~2007, covering (Semantic) Web, information retrieval, database and knowledge management:
\begin{itemize}
    \item conferences: WWW, WISE, APWeb, ISWC, ESWC, JIST, SIGIR, WSDM, ECIR, CIKM, SIGMOD, PODS, PVLDB, ICDE, EDBT, and
    \item journals: ACM Trans. Web, World Wide Web J., J. Web Semant., Semant. Web J., Int. J. Semant. Web Inf. Syst., ACM Trans. Inf. Syst., Inf. Process. Manag., Inf. Retr. J., Inf. Syst., IEEE Trans. Knowl. Data Eng., Data Knowl. Eng., Knowl. Inf. Syst., ACM Trans. Database Syst., VLDB J.
\end{itemize}

Second, we used the following query to search for papers in Google Scholar:
\begin{quote}
    ``entity summarization'' OR ``entity summarisation'' \,.
\end{quote}

We chose relevant papers from the collected papers based on whether their studied problem is compatible with our problem statement presented in Section~\ref{sect:ps}. Furthermore, we followed citations and references to consider additional papers that cite or are cited by the above relevant papers. With the defined search strategy and selection criterion, we are confident that the risk of introducing a researcher bias into the survey is low.

\subsection{Related Problems and Surveys}

There have been a few survey papers for related summarization problems including document summarization, graph summarization, and ontology summarization.

\paragraph{Document Summarization}

Document summarization has been studied for decades~\cite{ts1,ts2}. It is fundamentally different from entity summarization. Triples in an entity description are structured, whereas sentences in a document are unstructured text. However, some techniques for document summarization have inspired the development of entity summarizers. For example, RELIN~\cite{relin} which is an early entity summarizer uses a graph-based model which originates from the LexRank method~\cite{lexrank} for document summarization.

\paragraph{Graph Summarization}

Semantic data can be represented as a graph. Graph summarization aims to reveal patterns in the data. A graph summary is generally an abstract representation of the original graph. For example, it can be a single super-graph where super-vertices represent collections of vertices in the original graph; or it can be a set of frequent subgraph patterns. In the graph representation of semantic data, the description of an entity is represented as the neighborhood of the entity. Interesting patterns can hardly be mined from such a small star-shaped subgraph, and hence methods for graph summarization~\cite{gs1,gs2} are not suitable for entity summarization. Instead, entity summarizers usually adopt extractive solutions.

\paragraph{Ontology Summarization}

An ontology provides an explicit specification of a vocabulary for a shared domain. It is often used as a schema of semantic data. Methods for ontology summarization mainly represent an ontology as a graph, and then extract a subset of top-ranked terms and/or axioms based on graph centrality~\cite{os1,os2}. These methods are not suitable for entity summarization because their graph representations are specifically designed for schema-level ontologies rather than instance-level entity descriptions. However, graph centrality measures are universal and may apply to the graph representation of semantic data.

\subsection{Contribution of the Article}

The far-reaching application of entity summarization has led to fruitful research outcomes in recent years. However, there is a lack of comprehensive literature survey on this research topic. To the best of our knowledge, we provide the first technical review of existing entity summarizers. Instead of separately describing each summarizer, we identify their technical features and hierarchically organize the features to categorize the broad spectrum of research on entity summarization. Then we investigate various ways of combining multiple features to assemble a full entity summarizer. Our survey also addresses recent efforts on using deep neural networks for entity summarization, and covers known benchmarks and efforts for evaluation as well as a case study. Based on the review, we suggest directions for future work to overcome the limitations of existing research.

\subsection{Structure of the Article}

The rest of this article is organized as follows. Section~\ref{sect:features} reviews technical features for entity summarization. Section~\ref{sect:frameworks} reviews frameworks for feature combination and describes a set of representative entity summarizers. Section~\ref{sect:deep} reviews recent deep learning based entity summarizers. Section~\ref{sect:evaluation} reviews evaluation efforts for entity summarization. Finally, we suggest future directions in Section~\ref{sect:future}.
\section{Technical Features for Entity Summarization}
\label{sect:features}

\begin{figure}[!t]
	\centering
	\includegraphics[width=0.6\linewidth]{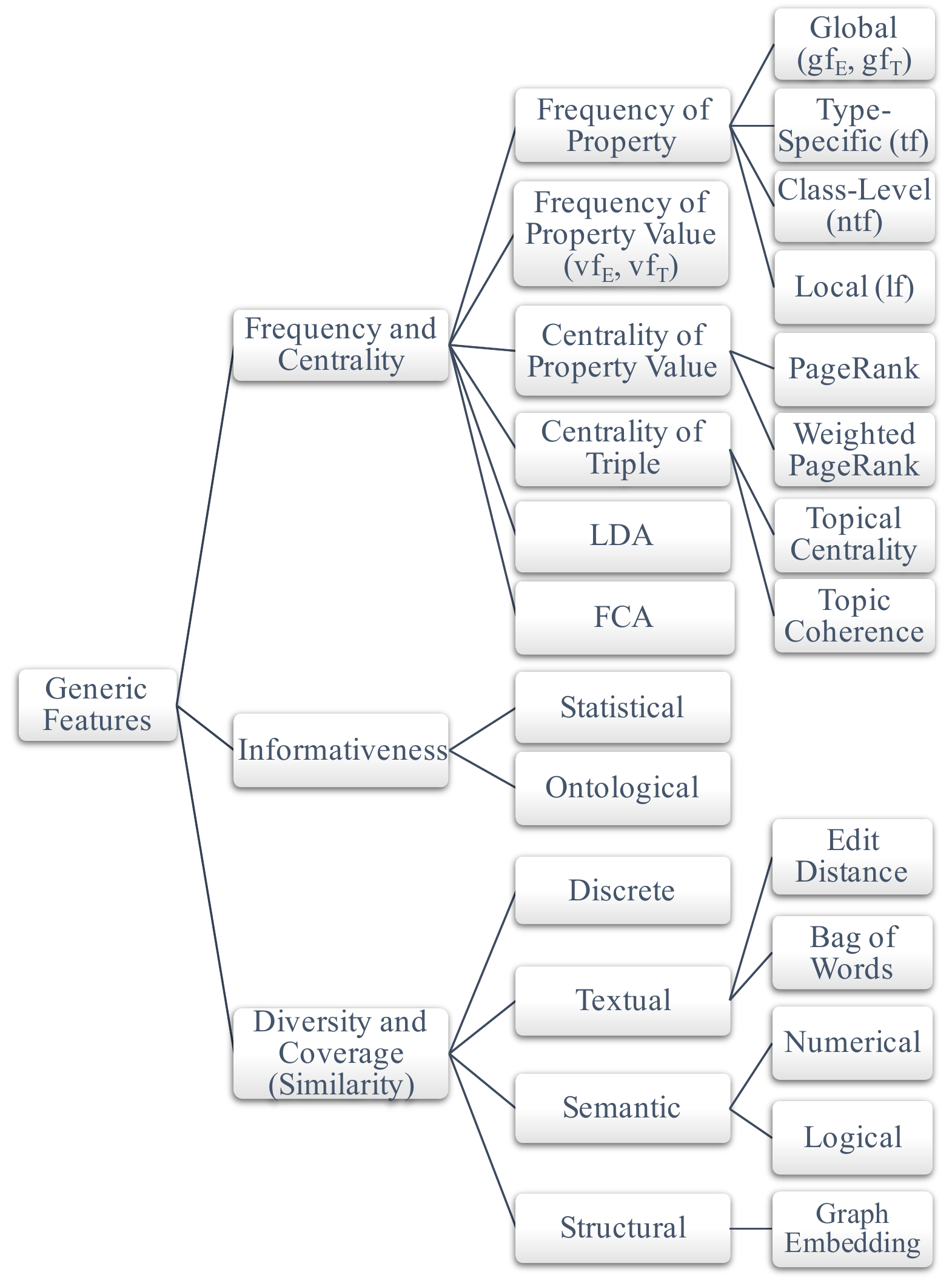}
	\caption{A hierarchy of generic features for entity summarization.}
	\label{fig:tech1}
\end{figure}

\begin{figure}[!t]
	\centering
	\includegraphics[width=0.6\linewidth]{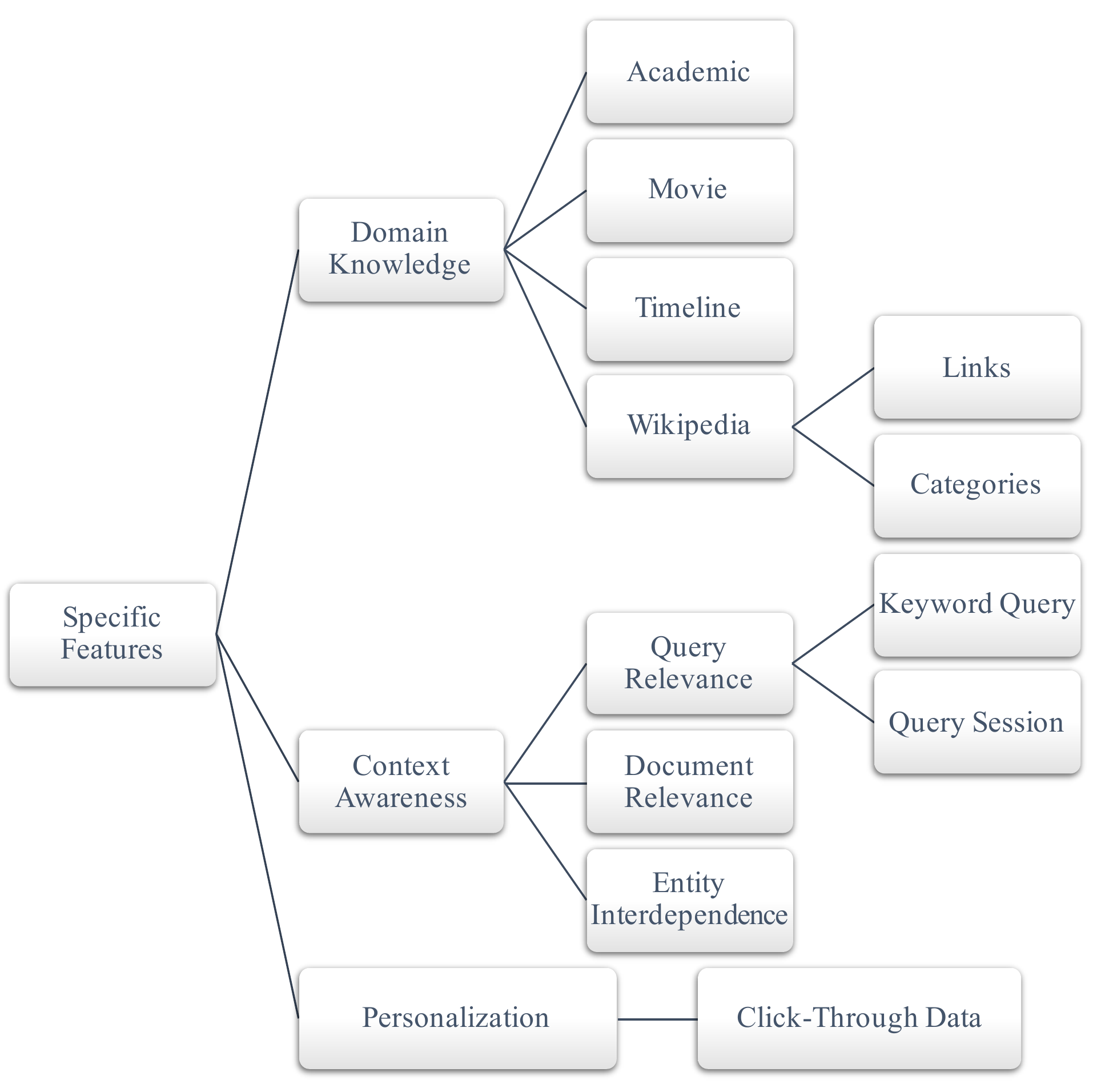}
	\caption{A hierarchy of specific features for entity summarization.}
	\label{fig:tech2}
\end{figure}

We broadly divide the technical features used in existing entity summarizers into generic features and specific features, and we organize them into hierarchies that are presented in Figure~\ref{fig:tech1} and Figure~\ref{fig:tech2}, respectively. Many of these features can find their counterparts in document summarization~\cite{ts1,ts2}, but they have been successfully adapted for semantic data. Table~\ref{table:algo} lists entity summarizers and the technical features they use.

\begin{table}[!t]
    \caption{Entity Summarizers (Sorted by Publication Date) and Their Technical Features (Blanks: Not Used)}
	\label{table:algo}
	\centering
    \resizebox{\columnwidth}{!}{
	\begin{tabular}{|l|p{3.9cm}|l|p{3.0cm}|p{3.2cm}|}
		\hline
		 & \multicolumn{3}{c|}{Generic Features} & Specific Features \\
		 \cline{2-5}
		 & Frequency and Centrality & Informativeness & Diversity and Coverage (Similarity) & Domains, Contexts, and Personalization \\
		\hline
		 Falcons~\cite{falcons} &  &  & bag-of-words & query relevance \\ 
		\hline
		 XRed~\cite{diff09} &  &  &  & entity interdependence \\ 
		\hline	
		 Zhang et al.~\cite{ibmcdl} & weighted PageRank &  &  & click-through data \\ 
		\hline
		RELIN~\cite{relin} & weighted PageRank & statistical &  & \\ 
		\hline
		 Thalhammer et al.~\cite{usewod} &  & statistical &  & movie domain \\ 
		\hline
		Yovisto~\cite{yovisto} & $\typefreq$ &  &  & academic domain, Wikipedia links \\ 
		\hline
		MMR-QSFS~\cite{sigir12} &  &  & 
		& query relevance \\ 
		\hline
		DIVERSUM~\cite{diversum} & $\localfreq$ &  & discrete & \\ 
		\hline
		SUMMARUM~\cite{summarum} & PageRank &  & discrete & \\ 
		\hline
		FACES~\cite{faces} & $\vfreq_\mathbb{T}$ & statistical & discrete, bag-of-words & \\ 
		\hline
		COMB~\cite{summel} & $\localfreq$ & statistical & edit-distance-like, numerical, logical & document relevance, entity interdependence \\ 
		\hline
		TimeMachine~\cite{timemachine} &  &  &  & timeline domain \\  
		\hline
	    C3D+P~\cite{c3dp} & $\localfreq$ & statistical & edit-distance-like, numerical, logical & entity interdependence \\ 
		\hline
		TRank\textsuperscript{++}~\cite{trank} & $\vfreq_\mathbb{E}$ & ontological &  & document relevance \\ 
		\hline
		FACES-E~\cite{facese} & $\vfreq_\mathbb{T}$ & statistical & discrete, bag-of-words & \\ 
		\hline
		CD~\cite{ensec1} &  & statistical & edit-distance-like, numerical, logical & \\ 
		\hline
	    Li et al.~\cite{ensec2} & $\globalfreq_\mathbb{E}$, $\localfreq$ &  &  & movie domain \\ 
		\hline
	    CES~\cite{waim} & weighted PageRank & statistical &  & session relevance \\ 
		\hline
		LinkSUM~\cite{linksum} & $\globalfreq_\mathbb{T}$, $\localfreq$, PageRank &  & discrete & Wikipedia links \\ 
		\hline
		Aemoo~\cite{aemoo} & $\typefreq$ &  &  & \\ 
		\hline
	    DynES~\cite{dynes} & $\globalfreq_\mathbb{E}$, $\globalfreq_\mathbb{T}$, $\numtypefreq$, $\vfreq_\mathbb{E}$, $\vfreq_\mathbb{T}$ &  &  & query relevance \\ 
		\hline
		REMES~\cite{ijcai17} & $\vfreq_\mathbb{E}$ & statistical & bag-of-words, structural & entity interdependence \\ 
		\hline
		Multi-EGS~\cite{multiegs} & $\typefreq$, $\numtypefreq$, $\vfreq_\mathbb{T}$ &  & edit-distance-like & Wikipedia categories \\ 
		\hline
		ES-LDA~\cite{eslda} & LDA &  &  & Wikipedia categories \\ 
		\hline
		ES-LDA$_{ext}$~\cite{esldaext} & LDA &  &  & \\ 
		\hline
		CTab~\cite{ctab} & $\globalfreq_\mathbb{E}$ &  & bag-of-words, numerical & entity interdependence \\ 
		\hline
 		BAFREC~\cite{bafrec} & $\globalfreq_\mathbb{T}$, $\vfreq_\mathbb{T}$ & ontological & bag-of-words &  \\ 
		\hline
 		KAFCA~\cite{kafca} & FCA & & &  \\ 
		\hline
 		MPSUM~\cite{mpsum} & LDA &  & discrete & Wikipedia categories \\ 
		\hline
	    Gottschalk et al.~\cite{eventkg} &  &  &  & timeline domain, Wikipedia links\\ 
		\hline
        VISION-KG~\cite{visionkg}
        & $\globalfreq_\mathbb{E}$, $\globalfreq_\mathbb{T}$, $\numtypefreq$, $\vfreq_\mathbb{E}$, $\vfreq_\mathbb{T}$ 
        & 
        & structural 
        & query relevance 
        \\
        \hline
		
	\end{tabular}
	}
\end{table}

\subsection{Generic Features}
\label{sect:generic}

Generic features apply to a wide range of domains, tasks, and users. We identify generic features from existing entity summarizers and group them into three categories: frequency/centrality, informativeness, and diversity/coverage.

\subsubsection{Frequency and Centrality}\label{sect:freq}

To generate a summary, an entity summarizer usually needs to measure the salience of each single triple for ranking. In the following we will first show measures based on frequency and centrality. Frequency is a common measure that has been widely used in document summarization~\cite{ts1,ts2}. Words that are frequently seen in a document are often topical words repeated by the author, and thus are usually (though not always) believed to be important. Graph centrality computes extended frequency and may be more effective and robust.

\textbf{Frequency of Property.}
To score an entity-property-value triple for entity summarization, \cite{summel,c3dp,ctab,ensec2,linksum,yovisto,diversum,aemoo,dynes,multiegs,bafrec,visionkg}~calculate the frequency of occurrence of the property over different scopes of data and interpret frequency in different ways.

In~\cite{ctab,ensec2,dynes,visionkg}, the \emph{global frequency} of a property~$p$ is calculated over all the entity descriptions in a dataset:
\begin{equation}\label{eq:gfe}
    \globalfreq_\mathbb{E}(p) = |\{ e' \in \mathbb{E} : \exists t \in \desc(e'), ~\prop(t)=p \}| \,.
\end{equation}
\begin{example}
    In Figure~\ref{fig:exampletable}, \texttt{worksAt} describes three entities (\texttt{Qu}, \texttt{Cheng}, and \texttt{Hu}) and hence $\globalfreq_\mathbb{E}(\texttt{worksAt})=3$; \texttt{directs} describes only one entity (\texttt{Qu}) and hence $\globalfreq_\mathbb{E}(\texttt{directs})=1$.
\end{example}
\noindent Another way of calculating global frequency is done at the triple level~\cite{linksum,dynes,bafrec,visionkg}:
\begin{equation}\label{eq:gft}
    \globalfreq_\mathbb{T}(p) = |\{ t \in \mathbb{T} : \prop(t)=p \}| \,.
\end{equation}
\noindent Its slight difference from~$\globalfreq_\mathbb{E}$ is that $\globalfreq_\mathbb{T}$ magnifies the frequency of multi-valued properties.
\begin{example}
    In Figure~\ref{fig:exampletable}, while $\globalfreq_\mathbb{E}(\texttt{knows})=\globalfreq_\mathbb{E}(\texttt{worksAt})=3$, \texttt{knows} is a multi-valued property and has a higher triple-level frequency $\globalfreq_\mathbb{T}(\texttt{knows})=5 > \globalfreq_\mathbb{T}(\texttt{worksAt})=3$.
\end{example}

In~\cite{yovisto,aemoo,multiegs}, frequency is calculated over only those entity descriptions of the same type. Priority is given to \emph{type-specific frequent properties} rather than generically frequent properties describing various types of entities. Formally, let $c \in \mathbb{C}$ be the type of the entity to be summarized, and let $\inst(c) \subseteq \mathbb{E}$ be the instances of~$c$, i.e., the set of entities having~$c$ as their type. The type-specific frequency of a property~$p$ is
\begin{equation}
    \typefreq(p) = |\{ e' \in \inst(c) : \exists t \in \desc(e'), ~\prop(t)=p \}| \,.
\end{equation}
\begin{example}
    In Figure~\ref{fig:exampletable}, \texttt{worksAt} describes only one entity (\texttt{Qu}) of type \texttt{Professor} and hence $\typefreq(\texttt{worksAt})=1$ w.r.t. $\inst(\texttt{Professor})=\{\texttt{Qu}\}$.
\end{example}

A related method is to calculate the number of types of entities that a property~$p$ describes~\cite{dynes,multiegs,visionkg}:
\begin{equation}
    \numtypefreq(p) = |\{ c' \in \mathbb{C} : \exists e' \in \inst(c'), ~t \in \desc(e'), ~\prop(t)=p \}| \,.
\end{equation}
\noindent It calculates frequency \emph{at the class level}.
\begin{example}
    In Figure~\ref{fig:exampletable}, \texttt{worksAt} describes two types of entities (\texttt{Person} and \texttt{Professor}) and hence $\numtypefreq(\texttt{worksAt})=2$.
\end{example}

In~\cite{summel,c3dp,ensec2,linksum,diversum}, the scope of calculation is narrowed to $\desc(e)$---the entity description to be summarized, to assess the \emph{local frequency} of a property~$p$:
\begin{equation}\label{eq:lf}
    \localfreq(p) = |\{ t \in \desc(e) : \prop(t)=p \}| \,.
\end{equation}
\noindent Again, (locally) multi-valued properties have higher frequency.
\begin{example}
    Inside the description of \texttt{Qu} in Figure~\ref{fig:exampletable}, \texttt{knows} describes \texttt{Qu} in two triples and hence $\localfreq(\texttt{knows})=2$; \texttt{worksAt} appears in only one triple and hence $\localfreq(\texttt{worksAt})=1$.
\end{example}

Whereas most of the above methods favor frequent properties, some prefer \emph{infrequent properties}, leading to seemingly contradictory conclusions. For example, in~\cite{ensec2}, both the most frequent and the most infrequent properties are prioritized. Similarly, in~\cite{linksum}, globally frequent but locally infrequent properties are preferred because they show (global) popularity and (local) exclusivity. Locally infrequent properties are also believed to be discriminating in~\cite{summel,c3dp}, and hence are useful for comparing entities. This contradicts the preference for locally frequent properties in~\cite{diversum}. In~\cite{multiegs}, properties that have higher type-specific frequency (i.e.,~$\typefreq$) but describe fewer types of entities (i.e.,~$\numtypefreq$) are ranked higher, which is conceptually similar to the TF-IDF scheme (short for term frequency---inverse document frequency) used in document summarization~\cite{ts1,ts2}. In Section~\ref{sect:inf} we will discuss infrequency from the angle of informativeness.

\textbf{Frequency of Property Value.}
In~\cite{dynes,visionkg}, the score of a triple depends on the frequency of occurrence of the property value~$v$ over all the entity descriptions in a dataset:
\begin{equation}
    \vfreq_\mathbb{E}(v) = |\{ e'\in E: \exists t \in \desc(e'), \val(t)=v \}| \,.
\end{equation}
\begin{example}
    In Figure~\ref{fig:exampletable}, \texttt{Jia} appears as a property value describing three entities (\texttt{Qu}, \texttt{Cheng}, and \texttt{Hu}) and hence $\vfreq_\mathbb{E}(\texttt{Jia})=3$; \texttt{Zhang} describes two entities (\texttt{Qu} and \texttt{Cheng}) and hence $\vfreq_\mathbb{E}(\texttt{Zhang})=2$.
\end{example}
\noindent A slightly different way of calculating frequency is done at the triple level~\cite{faces,facese,dynes,ijcai17,multiegs,bafrec,visionkg}:
\begin{equation}
    \vfreq_\mathbb{T}(v) = |\{ t \in \mathbb{T} : \val(t)=v \}| \,.
\end{equation}
\begin{example}
    In Figure~\ref{fig:exampletable}, while $\vfreq_\mathbb{E}(\texttt{Websoft})=\vfreq_\mathbb{E}(\texttt{Jia})=3$, \texttt{Websoft} has a higher triple-level frequency $\vfreq_\mathbb{T}(\texttt{Websoft})=4 > \vfreq_\mathbb{T}(\texttt{Jia})=3$.
\end{example} 

In particular, if the property is restricted to \texttt{TYPE} and the value is a class, the frequency of this class amounts to the number of its instances. In~\cite{trank}, this number is calculated over different scopes of entities, such as all the entities in a dataset or all the similar/related entities.


\textbf{Centrality of Property Value.}
In the graph representation of semantic data, the frequency of a property value which is a vertex in the graph is called its degree. Degree is one of the simplest \emph{graph centrality} measures, using only the neighborhood structure of a vertex. More generally, graph centrality assesses the extent to which a vertex is central in a graph. Among others, the well-known PageRank algorithm~\cite{pr} uses a model of a random surfer who continuously walks in the graph, either moving from a vertex to a random neighbor or jumping to a random vertex. The probability that the surfer is located at a vertex after a sufficiently large number of steps is defined as the PageRank centrality of the vertex. Compared with degree, PageRank is more powerful as it exploits the whole structure of a graph. PageRank has been used to score property values in entity summarization~\cite{linksum,summarum}.

In the original implementation of PageRank, the outgoing arcs of a vertex are uniformly weighted, and the random surfer is equally likely to move to any neighbor. However, arcs in semantic data can heterogeneously represent different types of relations having different semantics.
\begin{example}
    In Figure~\ref{fig:examplegraph}, the outgoing arcs of \texttt{Qu} represent three types of relations (\texttt{knows}, \texttt{worksAt}, and \texttt{directs}).
\end{example}
\noindent The random surfer may be more likely to move along certain arcs. Taking this into account, \cite{ibmcdl}~allows the random surfer to choose different arcs with different probabilities, which are defined in a domain-specific way.

\textbf{Centrality of Triple.}
The weighted PageRank model is also adopted by~\cite{relin}, which presents a new graph representation for entity descriptions inspired by LexRank~\cite{lexrank} for document summarization. The idea is to represent all the triples in an entity description as vertices of a complete graph. Every pair of vertices is connected by an arc weighted by the relatedness (i.e., similarity) of the two corresponding triples.
\begin{example}
    In Figure~\ref{fig:exampletable}, the description of \texttt{Qu} is represented by a complete graph containing eight vertices, each representing a triple in the description.
\end{example}
\noindent A triple is directly scored as a single unit by running PageRank on this graph, to assess its topical centrality. This weighting scheme is extended in~\cite{waim} to also consider topic coherence in a query context. Compared with the separation of property scoring and value scoring, such a joint model seems more suitable for the entity summarization task.

\textbf{LDA.}
Going beyond frequency, recent research has started to adapt more powerful statistical methods in information retrieval for entity summarization. Among others, Latent Dirichlet Allocation (LDA)~\cite{lda} is a generative statistical model, assuming that each document is a mixture of a small number of topics and each topic uses a small number of words frequently. In~\cite{eslda,esldaext,mpsum} where LDA is used for entity summarization, properties are treated as topics, and each property is a distribution over all the property values. With an LDA model learned from data, properties and/or their values are scored by their probabilities.

\textbf{FCA.}
In~\cite{kafca}, the Formal Concept Analysis (FCA) is performed to aggregate properties and values into a hierarchy. A triple is scored based on the depth of its elements in the hierarchy. The method implicitly gives preference to triples with infrequent properties and frequent values. Here, frequency is computed at the word level.

\subsubsection{Informativeness}\label{sect:inf}

Another group of features for measuring the salience of a triple is informativeness. This concept has been widely implemented in document summarization~\cite{ts1,ts2}, where a word appearing in fewer documents carries more information, and thus is more important. In the following we show two kinds of informativeness measures used in existing entity summarizers.

\textbf{Statistical Informativeness.}
A property-value pair $\langle p,v \rangle$ describing fewer entities in a dataset is considered more important~\cite{ijcai17,summel,c3dp,relin,usewod,waim,faces,facese,ensec1}. This statistical method has \emph{statistical and information theoretic explanations}. We can treat the occurrence of $\langle e,p,v \rangle$ in $\desc(e)$ as a probabilistic event. The informativeness of $\langle p,v \rangle$, i.e., the information content associated with the event of its occurrence, is measured by self-information, namely the negative logarithm of the probability of this event, where the probability is estimated using relative frequency observed in a dataset:
\begin{equation}\label{eq:si}
    \selfinfo(\langle p,v \rangle) = -\log \frac{|\{ e' \in \mathbb{E} : \langle e',p,v \rangle \in \desc(e') \}| }{|\mathbb{E}|} \,.
\end{equation}
\noindent A property-value pair that occurs less frequently will, once observed, carry a larger amount of information. Therefore, the self-information of a property-value pair shows its power of characterizing an entity.
\begin{example}
    In Figure~\ref{fig:exampletable}, $\langle\texttt{worksAt}, ~\texttt{Websoft}\rangle$ occurs in three out of the eight entity descriptions in the data and hence $\selfinfo(\langle\texttt{worksAt}, ~\texttt{Websoft}\rangle) = -\log \frac{3}{8}$; $\langle\texttt{directs}, ~\texttt{Websoft}\rangle$ occurs in only one entity description and hence $\selfinfo(\langle\texttt{directs}, ~\texttt{Websoft}\rangle) = -\log \frac{1}{8}$ representing higher informativeness.
\end{example}

Note that frequency, infrequency, and informativeness not necessarily conflict. They may be observed in different aspects of a property-value pair. For example, it is possible that an informative property-value pair has a locally infrequent property and has a globally frequent entity as the value.
\begin{example}
    In Figure~\ref{fig:exampletable}, $\langle\texttt{directs}, ~\texttt{Websoft}\rangle$ has high statistical informativeness. The local frequency of the property $\localfreq(\texttt{directs})=1$ is very low while the global frequency of the value $\vfreq_\mathbb{T}(\texttt{Websoft})=4$ is relatively high.
\end{example}

\textbf{Ontological Informativeness.}
If the property is restricted to \texttt{TYPE} and the value is a class, there is another way of measuring informativeness based on \emph{ontological semantics} in the schema of data, aka an ontology. Classes in an ontology typically form a subsumption hierarchy.
\begin{example}
    In Figure~\ref{fig:exampletable}, \texttt{Professor} can be a sub-class of \texttt{Person} in the schema.
\end{example}
\noindent A class deeper in the hierarchy has a more specific meaning than an upper-level one, and its occurrence carries more information~\cite{trank,bafrec}. Not surprisingly, such a class also has larger self-information because it has fewer instances. However, in practice, statistical informativeness and ontological informativeness are not equivalent due to the unbalance of the class hierarchy (hurting the accuracy of ontological informativeness) and/or the incompleteness of the data (hurting the accuracy of statistical informativeness).

\subsubsection{Diversity and Coverage}\label{sect:div}
Simply choosing a subset of the most salient triples may not generate a good summary because the information they provide may cover limited aspects of an entity, and on the other hand overlap with each other. Improving the diversity of a summary to more comprehensively cover the original information has been considered in document summarization~\cite{ts1,ts2} and also in entity summarization. The core task here is to measure the similarity between triples, and then avoid selecting very similar triples into a summary. In the following, we show various measures of similarity between two triples.

\textbf{Discrete Similarity.}
One simple method is to choose triples that have different properties~\cite{diversum,faces,facese,mpsum} or different values~\cite{linksum,summarum} for a summary.
This method essentially uses discrete similarity. For two properties (or two values) $i$ and $j$, their similarity is binary:
\begin{equation}
    \discr(i,j)=
    \begin{cases}
        1 & \text{if $i=j$,} \\
        0 & \text{if $i \neq j$.}
    \end{cases}
\end{equation}
\noindent That is, different properties and different values are individually separate. Implicit relationships between them are not explored.

\textbf{Textual Similarity.}
Properties and values are not just symbols but associated with natural language labels, from which textual similarity can be computed.

In~\cite{summel,c3dp,ensec1,multiegs}, \emph{edit-distance-like string metrics} are used to measure the similarity between the names of two properties and the similarity between the string forms of two property values.
\begin{example}
    In Figure~\ref{fig:exampletable}, \texttt{Websoft} and ``Director of Websoft'' are similar property values.
\end{example}

In~\cite{ijcai17,ctab,faces,facese,falcons,bafrec}, a string (e.g., ``Director of Websoft'') is represented as \emph{a bag of words} (e.g., ``director'', ``of'', ``websoft''), or a word vector. The similarity between two triples can be the cosine similarity between the two corresponding vectors~\cite{falcons}. To handle the sparseness of vectors and glean sense-level or higher-level abstract meanings of words, \cite{ijcai17,ctab,faces,facese}~use WordNet~\cite{wn} to expand words by including their synonyms or hypernyms.

\textbf{Semantic Similarity.}
Going beyond the superficial discrete and textual methods, semantic similarity aims to more deeply understand the meaning of data.

In~\cite{summel,c3dp,ctab,ensec1}, when comparing the string forms of two property values, if both of them represent numbers, their string similarity will be substituted by a special \emph{numerical similarity} which treats them as numbers instead of ordinary strings. In~\cite{summel,c3dp,ensec1}, the similarity between two numbers~$n_i$ and~$n_j$ is computed as follows, ranging from~-1 (dissimilar) to~1 (similar):
\begin{equation}
\begin{split}
\numsim(n_i,n_j)=
\begin{cases}
    1 & \text{if $n_i=n_j$,} \\
    -1& \text{if $n_i \neq n_j$ and $n_i n_j \leq 0$,} \\
    \frac{\min\{|n_i|,|n_j|\}}{\max\{|n_i|,|n_j|\}} & \text{if $n_i \neq n_j$ and $n_i n_j > 0$.}
\end{cases}
\end{split}
\end{equation}
\begin{example}
    ``99''~and~``100'' comprise different characters, but their numerical similarity $\numsim(99,100)=\frac{99}{100}$ is reasonably large.
\end{example}

In~\cite{summel,c3dp,ensec1}, they exploit ontological semantics to identify redundant triples via {logical reasoning}.
\begin{example}
    In the description of \texttt{Qu} in Figure~\ref{fig:exampletable}, $\langle\texttt{Qu}, ~\texttt{TYPE}, ~\texttt{Person}\rangle$ and $\langle\texttt{Qu}, ~\texttt{TYPE}, ~\texttt{Professor}\rangle$ are redundant and will not be selected together into an entity summary because \texttt{Professor} is a sub-class of \texttt{Person}.
\end{example}

\textbf{Structural Similarity.}
By representing semantic data as a graph, similarity can be computed based on the graph structure. Graph embedding techniques~\cite{DBLP:journals/tkde/CaiZC18} convert vertices into vectors in a low dimensional space where graph structural information and graph features are maximally preserved.

In~\cite{ijcai17}, the similarity between two property values is measured by the cosine angle between their embedding vectors generated by the RDF2Vec model~\cite{rdf2vec}.

In~\cite{visionkg}, embedding vectors for property values are generated by the SSP model~\cite{ssp}. For the set of property values in an entity summary, their embedding vectors are projected by locality-sensitive hashing into segmented real values where similar vectors will collide with high probability. Therefore, the number of projected segments represents diversity.

\subsection{Specific Features}

All the aforementioned features assess the salience of triples only by analyzing semantic data, thereby being universally usable. By contrast, the following features exploit external resources or consider external factors. They exhibit effectiveness for specific domains or specific tasks. These specific features use domain knowledge, be informed of contexts, or realize personalization. While some of these specific features are conceptually generalizable, e.g.,~for many domains we can exploit certain domain-specific knowledge, it would be labour-intensive to adapt existing concrete implementation to other domains and tasks, e.g.,~identifying and collecting useful knowledge for a given domain.

\subsubsection{Domain Knowledge}\label{sect:domainknowledge}

To summarize entities in a specific domain, it is possible and sometimes essential to obtain prior domain knowledge about the importance of certain triples. For example, \cite{yovisto}~summarizes the descriptions of academic videos, and gives priority to triples involving \texttt{Place} or \texttt{Event} entities because these classes are known to be important in the domain. In~\cite{ensec2}, to summarize the description of a film, the triple describing its most important actor is always selected into the summary. The importance of an actor is proportional to the number of films the actor stars in. In~\cite{usewod}, user ratings are leveraged to identify similar films by collaborative filtering. To summarize the description of a film, triples that also describe its most similar films are considered important. In~\cite{timemachine,eventkg}, a timeline of events is generated for an entity to show the most important milestones and relationships. During event selection, date relevance and temporal diversity are considered to avoid visual crowding along the time axis for easy user interaction.

In the description of an entity that has a Wikipedia page, a property value whose Wikipedia page is backlinked to the above page is considered important~\cite{linksum,yovisto,eventkg}. Wikipedia categories can be used to enrich semantic data for LDA-based methods~\cite{eslda,mpsum}. It is also possible to group entities by their Wikipedia categories from which an effective ranking of properties can be derived~\cite{multiegs}.

\subsubsection{Context Awareness}

For the same set of triples, some entity summarizers can produce different summaries depending on the context in which the generated summary is to be used.

In entity search engines, a \emph{keyword query} forms the context, and the relevance of a triple to the query is calculated for ranking and snippet generation~\cite{falcons,sigir12,dynes,visionkg}.
\begin{example}
    In Figure~\ref{fig:exampletable}, $\langle\texttt{Qu}, ~\texttt{directs}, ~\texttt{Websoft}\rangle$ is relevant to the query ``researchers at Websoft''.
\end{example}
\noindent This kind of context is extended from one single query to a \emph{query session} in~\cite{waim}.

In summary-assisted Web browsing, the \emph{content of a document} surrounding the mention of an entity forms the context. In~\cite{trank,summel}, preference is given to property values whose types are the same as or similar to the types of the other entities mentioned in the context.

In collective entity summarization~\cite{ijcai17,summel,c3dp,ctab,diff09}, summaries of multiple entity descriptions are jointly generated to help users compare or connect these entities. When summarizing each of these entity descriptions, their \emph{interdependence} forms the context. Their different values of the same property and/or their similar property-value pairs are prioritized.
\begin{example}
    In Figure~\ref{fig:exampletable}, when jointly summarizing the descriptions of \texttt{Qu} and \texttt{Cheng}, $\langle\texttt{Qu}, ~\texttt{worksAt}, ~\texttt{Websoft}\rangle$ and $\langle\texttt{Cheng}, ~\texttt{worksAt}, ~\texttt{Websoft}\rangle$ are likely to be selected because they provide a connection between the two entities.
\end{example}

\subsubsection{Personalization}\label{sect:personalization}
It is useful to generate personalized entity summaries to meet users' individual needs. For example, \cite{ibmcdl}~infers a user's preference from his or her \emph{clicks} when interacting with an application. More clicks on triples about a particular property indicate that the property is more important.
\section{Frameworks for Feature Combination and Representative Entity Summarizers}
\label{sect:frameworks}

A practical entity summarizer often relies on multiple features. However, different features may have conflicting objectives. In this section, we sketch out representative entity summarizers and show how they choose and combine multiple features.

\subsection{Simple Frameworks}

\textbf{Li et al.}
To integrate two or more features, one intuitive strategy is to separately rank triples based on each individual feature, and then take the union of their top-ranked results. For example, Li et al.~\cite{ensec2} compute the degree of specificness and the degree of generality for each property based on its local frequency~($\localfreq$) and global frequency~($\globalfreq_\mathbb{E}$), respectively. A generated entity summary consists of $z$~most specific properties and $k-z$~most general properties, where $0 \leq z \leq k$ is a parameter to tune.
\begin{example}
    In the description of \texttt{Qu} in Figure~\ref{fig:exampletable}, each triple is separately ranked by the specificness of its property which is proportional to~$\localfreq$, and by the generality of its property which is proportional to~$\globalfreq_\mathbb{E}$. Triples having the most specific properties such as \texttt{directs} and those having the most general properties such as \texttt{TYPE} will be selected into the summary.
\end{example}

\textbf{LinkSUM.}
Alternatively, multiple ranking criteria can be combined using an aggregate function, e.g., summation~\cite{linksum, visionkg}, multiplication~\cite{usewod,linksum}, or their mixture~\cite{multiegs}. For example, in LinkSUM~\cite{linksum}, properties and property values are scored separately. The score of a property~$p$ is given by the product of its three features:
\begin{equation}
\label{eq:linksum-p}
    \frq(p) * \exc(p) * \dsc(p) \,,
\end{equation}
\noindent where $\frq$~(short for frequency) represents global frequency~($\globalfreq_\mathbb{T}$), $\exc$~(short for exclusivity) represents the reciprocal of local frequency~($\localfreq$), and $\dsc$~(short for description) represents the number of annotations of~$p$, including the labels, domains, and ranges of~$p$. The score of a property value~$v$ is a linear combination of its normalized PageRank score and backlink~($\bl$) score:
\begin{equation}
\label{eq:linksum-v}
    \alpha \cdot \frac{\pagerank(v)}{\max_{v' \in \mathbb{E}}{\pagerank(v')}} + (1-\alpha) \cdot \bl(v) \,,
\end{equation}
\noindent where $\alpha \in [0,1]$ is a parameter to tune, and $\bl(v)=1$ if $v$~is backlinked to the entity to be summarized, otherwise $\bl(v)=0$. LinkSUM selects property values having the highest scores. When a property value appears in more than one triple, the property having the highest score is selected.
\begin{example}
    In the description of \texttt{Qu} in Figure~\ref{fig:exampletable}, for each triple its property and its value are scored by Eq.~(\ref{eq:linksum-p}) and Eq.~(\ref{eq:linksum-v}), respectively. Triples are ranked by the scores of their property values. Ties are broken by the scores of the properties; for each unique value (e.g.,~\texttt{Websoft}), only one triple having the top-ranked property (e.g.,~\texttt{worksAt}) is preserved.
\end{example}

\textbf{Remarks.}
Despite simplicity, the above frameworks are not suitable for incorporating diversity and coverage features discussed in Section~\ref{sect:div}.

\subsection{Random Surfer Model}

\textbf{RELIN.}
In RELIN~\cite{relin}, a weighted PageRank model is used to rank triples by computing their PageRank centrality in a graph where vertices represent triples to be ranked and are pairwise adjacent to each other. In PageRank, a random surfer at each step either moves from a vertex (i.e., a triple) to a random neighbor or jumps to a random vertex. The two types of actions can have different probabilities, which are exploited in RELIN to represent informativeness and relatedness. Specifically, the probability that a random surfer jumps to a random triple~$t$, denoted by~$p_J(t)$, is proportional to the statistical informativeness~($\selfinfo$) of~$t$:
\begin{equation}
    p_J(t) = \selfinfo(\langle \prop(t),\val(t) \rangle) \,.
\end{equation}
\noindent The probability that a random surfer moves from a triple~$t'$ to a random neighbor~$t$, denoted by~$p_M(t',t)$, is proportional to the topical relatedness between~$t'$ and~$t$. Topical relatedness is computed using pointwise mutual information (PMI), a measure of association used in information theory:
\begin{equation}
    p_M(t',t) = \sqrt{\pmi(\prop(t'),\prop(t)) \cdot \pmi(\val(t'),\val(t))} \,.
\end{equation}
\noindent To compute~$\pmi$, for two properties their names are used, and for two property values their string forms are used, denoted by~$s'$ and~$s$. RELIN employs a Web search engine to estimate probabilities in PMI computation:
\begin{equation}
    \pmi(s',s) = \frac{\frac{\hits(s';s)}{N}}{\frac{\hits(s')}{N} \cdot \frac{\hits(s)}{N}} \,,
\end{equation}
\noindent where $\hits$~represents the number of results returned by the Web search engine for a query, and $N$~is a large integer for normalization. Finally, the PageRank centrality of triple~$t$, i.e.,~the probability that a random surfer arrives at~$t$, is iteratively defined as
\begin{equation}
    \pr(t) = (1-d) \cdot p_J(t) + d \cdot \sum_{t' \in \nb(t)}{\pr(t') \cdot p_M(t',t)} \,,
\end{equation}
\noindent where $d \in [0,1]$~is a damping factor, and $\nb(t)$ denotes the neighbors of~$t$.
\begin{example}
    In the description of \texttt{Qu} in Figure~\ref{fig:exampletable}, the random surfer is more likely to jump to triples such as $\langle\texttt{Qu}, ~\texttt{directs}, ~\texttt{Websoft}\rangle$ which has high statistical informativeness, and is also more likely to move to this triple from some other triples such as $\langle\texttt{Qu}, ~\texttt{worksAt}, ~\texttt{Websoft}\rangle$ due to the high PMI between their (identical) property values.
\end{example}

\textbf{CES.}
The above weighted PageRank model can integrate more than two ranking features. The probability of a move or a jump can be defined as a linear combination of multiple ranking functions. For example, CES~\cite{waim} extends RELIN by adding session relevance to the model. Session relevance is the relevance of a triple to a query session. In CES, probability is a linear combination of statistical informativeness, topical relatedness, and session relevance.

\textbf{Remarks.}
PageRank and other random surfer models are suitable for integrating centrality-based and importance-style features. However, one of their shortcomings is that diversity and coverage features cannot be naturally modeled.

\subsection{Similarity-based Grouping}

\textbf{DIVERSUM.}
To explicitly support diversity and coverage, DIVERSUM~\cite{diversum} disallows selected triples to have the same property. In other words, triples are grouped by their properties. Only one triple in each group can be selected. To rank triples in different groups, the local frequency~($\localfreq$) of their properties is compared.
\begin{example}
    In the description of \texttt{Qu} in Figure~\ref{fig:exampletable}, the triples about \texttt{TYPE} and \texttt{knows} are prioritized due to their relatively high~$\localfreq$. However, for each of these properties, only one triple having this property will be selected into the summary.
\end{example}

\textbf{FACES(-E).}
A similar framework is adopted by FACES~\cite{faces} and its extension FACES-E~\cite{facese}. They represent a triple as a bag of words expanded with hypernyms in WordNet, and group triples by an incremental and hierarchical text clustering algorithm. The incremental nature of the approach enables it to work even in a streaming data environment where the total number of data points (i.e., triples) are not known a priori. Within each group, a triple~$t$ is scored by the product of its statistical informativeness~($\selfinfo$) and the frequency of its property value~($\vfreq_\mathbb{T}$):
\begin{equation}
   \selfinfo(\langle \prop(t),\val(t) \rangle) \cdot \vfreq_\mathbb{T}(\val(t)) \,.
\end{equation}
\noindent If the number of groups is larger than the number of triples to select (i.e.,~$k$), at most one top-ranked triple will be selected from each group. Otherwise, at least one top-ranked triple will be selected from each group. A similar strategy is adopted by \textbf{MPSUM}~\cite{mpsum}.
\begin{example}
    In the description of \texttt{Qu} in Figure~\ref{fig:exampletable}, the three triples where ``Websoft'' is mentioned may form a group where the top-ranked triple is $\langle\texttt{Qu}, ~\texttt{directs}, ~\texttt{Websoft}\rangle$ because it has the highest statistical informativeness and also the highest frequency of property value.
\end{example}

\textbf{Remarks.}
Although grouping-based frameworks support diversity and coverage features, they assume that the similarity between triples is an equivalence relation and induces a partition of triples. However, similarity is not necessarily a binary function but is generally a numerical function. A strict partitioning of triples may be inflexible and inaccurate.

\subsection{MMR-like Re-Ranking}

\textbf{MMR-QSFS.}
MMR~\cite{mmr} (short for Maximal Marginal Relevance) is an information retrieval framework that improves the diversity of the results by selecting items iteratively. In each iteration, the item to select is the candidate that maximizes its quality score and minimizes its similarity with the items already selected in previous iterations. Candidates are re-ranked in each iteration. MMR-QSFS~\cite{sigir12} adopts this framework to generate diversified query-relevant entity summaries. Specifically, let~$S$ be the current entity summary consisting of the triples selected in previous iterations. Initially $S$~is empty. In each iteration, the MMR score of each candidate triple~$t$ is given by
\begin{equation}
    \mmr(t) = \lambda \cdot \qr(t) - (1-\lambda) \cdot \max_{t' \in S}{\simi(t,t')} \,,
\end{equation}
\noindent where $\qr(t)$ is the relevance of~$t$ to the query, $\simi(t,t')$ is the similarity between two triples~$t$ and~$t'$, and $\lambda \in [0,1]$~is a parameter to tune.
\begin{example}
    In the description of \texttt{Qu} in Figure~\ref{fig:exampletable}, assume $\langle\texttt{Qu}, ~\texttt{worksAt}, ~\texttt{Websoft}\rangle$ is the triple selected in the first iteration. Then for the next iteration, the MMR score of the candidate triple $\langle\texttt{Qu}, ~\texttt{directs}, ~\texttt{Websoft}\rangle$ will be reduced because it is very similar to a selected triple.
\end{example}

\textbf{Remarks.}
This framework achieves a trade-off between importance and diversity when selecting triples. Compared with grouping, MMR is more flexible and fully exploits the numerical values of similarity. More generally, this trade-off can be reformulated as a linear combination of importance and diversity over all the triples in a summary, as we will see in Section~\ref{sect:combinatorial}. From this point of view---treating the linear combination as an objective function to optimize, MMR is actually a greedy algorithm which may not produce an optimum solution.

\subsection{Combinatorial Optimization}\label{sect:combinatorial}

\textbf{CD.}
To overcome the sub-optimality of MMR, it would be straightforward to directly model entity summarization as a combinatorial optimization problem. In CD~\cite{ensec1}, for each triple $t_i \in \desc(e)$, let~$x_i$ be a binary variable representing whether $t_i$~is selected into the summary ($x_i=1$) or not ($x_i=0$). The problem is to
\begin{equation}\label{eq:cd}
\begin{split}
    \text{maximize } & \sum_{i=1}^{|\desc(e)|}{\sum_{j=i}^{|\desc(e)|}{p_{ij}x_ix_j}} \\
    \text{subject to } & \sum_{i=1}^{|\desc(e)|}{x_i} \leq k \,,\\
    & x_i \in \{0,1\} \text{ for } i=1,\ldots,|\desc(e)| \,,\\
    \text{where } & p_{ij} =
    \begin{cases}
        \gamma \cdot \selfinfo(\langle \prop(t_i),\val(t_i) \rangle) & \text{if $i=j$} \,,\\
        \delta \cdot (-\simi(t_i,t_j)) & \text{if $i \neq j$} \,,
    \end{cases}
\end{split}
\end{equation}
\noindent where $\selfinfo(\langle \prop(t_i),\val(t_i) \rangle)$ is the statistical informativeness of~$t_i$, $\simi(t_i,t_j)$ is the similarity between~$t_i$ and~$t_j$ which integrates string similarity, numerical similarity, and logical reasoning, and $\gamma,\delta > 0$ are parameters to tune. The objective function is to maximize the statistical informativeness of selected triples while minimizing their similarity to improve diversity. This problem is an instance of the quadratic knapsack problem, which has effective heuristic algorithms~\cite{grasp}.
\begin{example}
    In Figure~\ref{fig:exampletable}, when summarizing the description of \texttt{Qu}, once $\langle\texttt{Qu}, ~\texttt{directs}, ~\texttt{Websoft}\rangle$ has been greedily selected into a summary due to its high statistical informativeness, it will be unlikely to select $\langle\texttt{Qu}, ~\texttt{worksAt}, ~\texttt{Websoft}\rangle$ into the same summary because it is very similar to a selected triple.
\end{example}

\textbf{C3D+P.}
The quadratic knapsack model is also adopted in C3D+P~\cite{c3dp}, where the goal is to generate a summary for comparing a pair of entities. The objective function considers not only statistical informativeness and intra-entity triple similarity as in Eq.~(\ref{eq:cd}), but also inter-entity triple similarity and dissimilarity for comparing entities.
\begin{example}
    In Figure~\ref{fig:exampletable}, when jointly summarizing the descriptions of \texttt{Qu} and \texttt{Cheng} for comparing them, their common property-value $\langle\texttt{knows}, ~\texttt{Zhang}\rangle$ will be selected into their summaries because it informatively characterizes a commonality between them.
\end{example}

\textbf{REMES.}
An extended quadratic multidimensional knapsack model is used in REMES~\cite{ijcai17}, where the goal is to generate a summary to characterize the relatedness among a set of entities~$E$. For each entity $e_i \in E$ and each triple $t_a \in \desc(e_i)$, let~$x_{i,a}$ be a binary variable representing whether $t_a$~is selected into the summary of~$e_i$ ($x_{i,a}=1$) or not ($x_{i,a}=0$). The problem is to
\begin{equation}\label{eq:remes}
\begin{split}
    \text{maximize } & \sum_{i=1}^{|E|}{\sum_{j=i}^{|E|}{\sum_{a=1}^{|\desc(e_i)|}{\sum_{b=1}^{|\desc(e_j)|}{p_{i,a,j,b} \cdot x_{i,a} \cdot x_{j,b}}}}} \\
    \text{subject to } & \sum_{a=1}^{|\desc(e_i)|}{x_{i,a}} \leq k \text{ for } i=1,\ldots,|E| \,,\\
    & x_{i,a} \in \{0,1\} \text{ for } i=1,\ldots,|E| \text{ and } a=1,\ldots,|\desc(e_i)| \,,\\
    \text{where } & p_{i,a,j,b} =
    \begin{cases}
        \alpha \cdot \selfinfo(\langle \prop(t_a),\val(t_a) \rangle) \cdot \vfreq_\mathbb{T}(\val(t_a)) & \text{if $i=j$ and $a=b$} \,,\\
        \beta \cdot (-\rel(t_a,t_b)) & \text{if $i=j$ and $a \neq b$} \,,\\
        \gamma \cdot \rel(t_a,t_b) & \text{if $i \neq j$} \,,
    \end{cases}
\end{split}
\end{equation}
\noindent where $\selfinfo(\langle \prop(t_a),\val(t_a) \rangle)$~is the statistical informativeness of~$t_a$, $\vfreq_\mathbb{T}(\val(t_a))$~is the frequency of the property value of~$t_a$, $\rel(t_a,t_b)$~is the relatedness between~$t_a$ and~$t_b$, and $\alpha,\beta,\gamma > 0$ are parameters to tune. The objective function is to maximize the statistical informativeness and property value frequency of selected triples, minimize the relatedness between selected intra-entity triples to improve diversity, and maximize the relatedness between selected inter-entity triples to characterize relatedness between entities. Relatedness is computed using both the overlap between two bags of words expanded with WordNet and the cosine angle between two graph embedding vectors.
\begin{example}
    In Figure~\ref{fig:exampletable}, when jointly summarizing the descriptions of \texttt{Qu}, \texttt{Cheng}, and \texttt{Hu} for showing their relatedness, their common property-value $\langle\texttt{worksAt}, ~\texttt{Websoft}\rangle$ will be selected into their summaries.
\end{example}

\textbf{COMB.}
The quadratic multidimensional knapsack model is also adopted in COMB~\cite{summel}, where the goal is to generate a summary for distinguishing between a set of candidate entities that may be mentioned in a document. The objective function considers not only statistical informativeness and triple dissimilarity but also the relevance of these entities to the document content. Entities and the document content are all represented as vectors in a space where each dimension represents a class. Weights in vectors are computed in a way similar to the well-known term frequency---inverse document frequency in information retrieval. The weight of a class is proportional to its number of instances in the document but inversely proportional to its total number of instances.

\textbf{Remarks.}
In addition to knapsack-like modeling, \cite{ctab,diff09,timemachine}~develop formulations of task-specific combinatorial optimization problems. Compared with MMR, combinatorial optimization provides a more principled way of formulating the problem of entity summarization. However, their difference is mainly theoretical rather than practical, because the formulated combinatorial optimization problems are often NP-hard and are solved by heuristic or greedy algorithms, which are conceptually similar to MMR.

\subsection{Learning to Rank}\label{sect:ltr}
From the view of machine learning, all the above frameworks are unsupervised. When labeled data is available for training, it is possible to consider a supervised learning framework. For example, for each triple we can define a feature vector where each feature is given by one of the above-mentioned technical features. Entity summarization is then modeled as a learning-to-rank problem, which can be solved by decision tree and linear regression~\cite{trank}, support vector machine~\cite{eventkg}, gradient tree boosting~\cite{sigir12,dynes}, or more sophisticated learners.

Supervised learning has so far not been widely used for entity summarization due to the shortage of labeled data. Existing attempts~\cite{trank,sigir12,dynes,eventkg} mainly focus on specific applications where labeled data is available, such as user clicks for the generation of query-relevant entity summaries. In Section~\ref{sect:deep}, we will detail two recent entity summarizers using deep neural networks.
\section{Deep Neural Networks for Entity Summarization}\label{sect:deep}

The success of deep learning in tasks like speech recognition, image classification, and natural language processing suggests a promising direction for entity summarization research. Recently, a few efforts~\cite{esa, deeplens} have been made to apply deep neural networks to the entity summarization task. The idea of these entity summarizers is to supervisedly learn a scoring function for triple ranking, where the scoring function is modeled as a parameterized deep neural network.

\textbf{ESA.}
The first entity summarizer using deep neural networks is ESA~\cite{esa}. In the first step, each triple in~$\desc(e)$ is represented by a vector concatenating the embedding vectors of its property and value. The embedding vector of the property is learned using a word embedding technique~\cite{DBLP:journals/jmlr/BengioDVJ03}. The embedding vector of the value is derived from a pre-trained TransE model~\cite{transe} which encodes structural features. In the second step, the vector representations of all the triples in~$\desc(e)$ are fed as a sequence into a BiLSTM network, which outputs a vector representation~$\mathbf{h}$ for~$\desc(e)$. Based on this representation, in the last step, a supervised attention mechanism is used to score each triple in~$\desc(e)$ by the dot product of its vector representation and~$\mathbf{h}$. These dot products generate the final attention vector after using softmax function. The learning task here is to minimize the cross-entropy loss between the generated attention vector and the attention vector derived from labeled entity summaries.

\textbf{DeepLENS.}
One shortcoming of ESA is, due to the use of BiLSTM for encoding a set as a sequence, the output is sensitive to the chosen order. DeepLENS~\cite{deeplens} overcomes this problem with a different network. In the first step, both the property and the value in each triple are represented by their pre-trained word embedding vectors, which in turn are concatenated to represent the triple. Different from ESA, graph structure is completely ignored in DeepLENS. In the second step, each candidate triple $t \in \desc(e)$ is encoded by a multi-layer perceptron (MLP) which outputs a vector~$\mathbf{h}$. All the triples in~$\desc(e)$ are encoded by another MLP and then are summed and weighted using attention from~$\mathbf{h}$ to generate the vector representation of~$\desc(e)$ denoted by~$\mathbf{d}$. This representation is permutation invariant to the order of the triples in~$\desc(e)$. In the last step, $\mathbf{h}$~and~$\mathbf{d}$ are concatenated and fed into a MLP to generate the score of~$t$. The learning task here is to minimize the mean squared error loss between the generated scores and the scores derived from labeled entity summaries.
 
\textbf{Remarks.}
Compared with traditional entity summarizers reviewed in Section~\ref{sect:frameworks} which choose and combine multiple features, deep learning based entity summarizers avoid manual feature engineering. However, they rely on a large number of labeled entity summaries for training. Manually labeling entity summaries is expensive.
\section{Evaluation of Entity Summarization}
\label{sect:evaluation}

Methods for evaluating entity summarization are broadly divided into intrinsic methods and extrinsic methods.

Intrinsic evaluation directly measures the quality of a machine-generated summary by comparing it with a human-made ground-truth summary. Intrinsic evaluation is popular because it is relatively easy to perform and the results are reproducible. In this section, we present metrics used in intrinsic evaluation, describe known benchmarks offering ground-truth summaries created for evaluation, and illustrate some evaluation results.

Extrinsic evaluation indirectly measures the quality of a machine-generated summary by applying it in a downstream task and measuring users' effectiveness and efficiency in completing the task based on the summary. Extrinsic methods are usually adopted to evaluate task-specific entity summarizers. In this section we will review extrinsic evaluation reported in the literature.
However, extrinsic evaluation is difficult to replicate as human users are involved.

\subsection{Intrinsic Evaluation}

\subsubsection{Evaluation Metrics}

Let~$S_m$ be a machine-generated entity summary. Let~$S_h$ be a human-made ground-truth summary. Intrinsic evaluation compares~$S_m$ with~$S_h$ and scores~$S_m$ based on the extent to which $S_m$~is similar to~$S_h$. When there are multiple ground-truth summaries made by different human experts for an entity, the mean score of~$S_m$ is calculated.

Two similarity metrics have been used in the literature: quality and F1.

\textbf{Quality Score.}
Entity summarizers take a size constraint~$k$ as input, to bound the number of triples in a summary. In intrinsic evaluation, $k$~is commonly set to~$|S_h|$, and hence we usually have $|S_m|=|S_h|$. In~\cite{relin,waim,faces,facese,linksum}, the quality score of~$S_m$ is computed based on its overlap with~$S_h$:
\begin{equation}
    \text{Quality} = |S_m \cap S_h| \,.
\end{equation}

One shortcoming of this simple metric is the lack of normalization. Therefore, the results in different settings of~$k$ are not comparable. Besides, this metric may not be fair when $|S_m| \neq |S_h|$.

\textbf{F1 Score.}
To overcome the above shortcomings, recent evaluation efforts use standard information retrieval metrics:
\begin{equation}
    \text{Precision} = \frac{|S_m \cap S_h|}{|S_m|} \,,\quad
    \text{Recall} = \frac{|S_m \cap S_h|}{|S_h|} \,,\quad
    \text{F1} = \frac{2 \cdot \text{Precision} \cdot \text{Recall}}{\text{Precision} + \text{Recall}} \,.
\end{equation}
\noindent The results of Precision, Recall, and F1 score are in the range of $[0,1]$.

Note that we will trivially have Precision$=$Recall$=$F1 if $|S_m|=|S_h|$. However, even if we set $k=|S_h|$, some entity summarizers may output less than $k$~triples, i.e.,~$|S_m|<|S_h|$. For example, DIVERSUM~\cite{diversum} disallows selected triples to have the same property. It is possible that an entity description contains less than $k$~distinct properties and hence DIVERSUM has to output less than $k$~triples. In this case, Precision$\neq$Recall, and one should rely on F1 score.

\textbf{Ranking-based Metrics.}
As mentioned in Section~\ref{sect:ps}, some methods formulate and solve entity summarization as a triple ranking problem. Given an input entity description~$\desc(e)$, they output a ranking of the triples in~$\desc(e)$. To evaluate this ranking, we can treat the $k$~top-ranked triples as a machine-generated summary and evaluate it using the above-mentioned set-based metrics such as Precision. That amounts to calculating Precision at~$k$, which is a popular evaluation metric in information retrieval. We can also use other information retrieval metrics such as Mean Average Precision (MAP) and Normalized Discounted Cumulative Gain (NDCG) to directly evaluate the entire ranking. We will not detail these metrics as they have not been widely used in entity summarization research. 

\begin{table}
	\caption{Benchmarks for Intrinsic Evaluation of Entity Summarization}\label{table:datasets}
	\centering
    \resizebox{\columnwidth}{!}{
	\begin{tabular}{|l|p{2cm}|p{1.6cm}|p{8cm}|}
		\hline
		 & Entities & Number of Entities & Entity Types \\
		\hline
		Benchmark for Evaluating RELIN~\cite{relin} & DBpedia & 149 & not specified \\
		\hline
		WhoKnows?Movies!~\cite{game}\footnotemark[1] & Freebase & 60 & movie \\
		\hline
		Benchmark for Evaluating DIVERSUM~\cite{diversum} & IMDb & 20 & actor \\
		\hline
		Langer et al.~\cite{assign} & DBpedia & 14 & theoretical physicist, Grammy winner, male ``best actor'', Academy Award winner, super heavyweight boxer, U.S. president, etc. \\
		\hline
		Benchmark for Evaluating FACES~\cite{faces}\footnotemark[2] & DBpedia & 50 & politician, actor, scientist, song, film, country, city, river, company, game, etc. \\
		\hline
		FRanCo~\cite{franco} & DBpedia & 265 & from 189~classes \\
		\hline
		Benchmark for Evaluating FACES-E~\cite{facese}\footnotemark[2] & DBpedia & 80 & not specified \\
		\hline
ESBM v1.2~\cite{esbm}\footnotemark[3] & DBpedia, LinkedMDB & 175 & agent, event, location, species, work, film, person \\
		\hline
	\end{tabular}
	}
\end{table}
\footnotetext[1]{\url{http://yovisto.com/labs/iswc2012}}
\footnotetext[2]{\url{http://wiki.knoesis.org/index.php/FACES}}
\footnotetext[3]{\url{https://w3id.org/esbm}}

\subsubsection{Benchmarks}
\label{sect:datasets}

Table~\ref{table:datasets} presents known efforts in creating benchmarks offering human-made ground-truth entity summaries or triple scores for intrinsic evaluation. All these ground truths are generic, i.e.,~they are created not for any particular task or any individual user, but for general purposes. At the time of writing this article, only four of these benchmarks for intrinsic evaluation of entity summarization are online accessible. Their hyperlinks are included in Table~\ref{table:datasets} as footnotes. Most of these efforts use entities in DBpedia which contains RDF data extracted from Wikipedia. Some use LinkedMDB or IMDb, both of which contain semantic data for the movie domain.

The authors of RELIN~\cite{relin} invited 24~human experts to independently create ground-truth summaries for 149~entities randomly selected from DBpedia. Each entity description contained 20--40~triples and was summarized by an average of 4.43~participants. A participant created two summaries: one containing 5~selected triples and the other containing 10~triples.

The authors of FACES~\cite{faces} and FACES-E~\cite{facese} followed a similar procedure. To evaluate FACES, 15~human experts were invited to independently create ground-truth summaries for 50~entities in DBpedia. Each entity was described by at least 17~distinct properties and an average of 44~triples, and was summarized by at least 7~participants. To evaluate FACES-E, 17~human experts were invited to summarize 80~entities. Each entity was summarized by at least 4~participants. These benchmarks are available online.

Langer et al.~\cite{assign} and FRanCo~\cite{franco} also used entities in DBpedia. Instead of entity summaries, they collected graded importance score of each triple. In~\cite{assign}, 10~human experts were invited to independently label the triples in 14~entity descriptions as highly relevant, relevant, less relevant, or irrelevant. In~\cite{franco}, 265~entities were processed by at least 5~participants. An entity description contained 10--150 triples with a mean value of~48. Each participant selected the importance of each fact in an entity description on the Likert scale, scoring it from~1 to~5.

Whereas all the above efforts used DBpedia, the authors of DIVERSUM~\cite{diversum} selected 20~entities from IMDb. They did not collect ground-truth summaries from human experts but extracted from the Wikipedia info-box concerning each entity, which was edited by a large population of editors and could be viewed as a high-quality entity summary.

The WhoKnows?Movies! game~\cite{game} was designed to collect ground-truth importance of triples. This online quiz game attracted 217~players who answered 8,308~questions about 2,829~triples in the descriptions of 60~movies taken from Freebase. A question was generated out of a triple asking for movies having a given property-value pair, e.g.,~an actor. The ratio of correctly answered questions based on a triple could be used as an indicator of the importance of the triple. This benchmark is available online.

The Entity Summarization Benchmark (ESBM)~v1.2~\cite{esbm} used entities from DBpedia and LinkedMDB. Ground-truth summaries for 175~entities were collected from 30~independent human experts. Each entity description contained at least 20~triples, and was summarized by 6~participants. A participant created two summaries: one containing 5~selected triples and the other containing 10~triples. At the time of writing this article, ESBM~v1.2 is the largest available benchmark for intrinsic evaluation of entity summarization. It has been used to evaluate a relatively large number of existing entity summarizers.

\subsubsection{Evaluation Results}\label{sect:intrinsic-results}

Most benchmarks described in Section~\ref{sect:datasets} have been used to evaluate none or only a few entity summarizers. ESBM~v1.2 is an exception, which was published together with the evaluation results of nine unsupervised entity summarizers: RELIN~\cite{relin}, DIVERSUM~\cite{diversum}, FACES~\cite{faces}, FACES-E~\cite{facese}, CD~\cite{ensec1}, LinkSUM~\cite{linksum}, BAFREC~\cite{bafrec}, KAFCA~\cite{kafca}, and MPSUM~\cite{mpsum}. In~\cite{deeplens}, the evaluation results of two deep learning based entity summarizers on ESBM~v1.2 were reported: ESA~\cite{esa} and DeepLENS~\cite{deeplens}. Here we will not repeat these detailed evaluation results but we highlight the main conclusion: DeepLENS outperforms all the competitors on ESBM~v1.2 but still, there is a large gap between DeepLENS and ORACLE which represents the best possible performance that can be achieved. For example, when $k=5$, for DBpedia entities the mean F1 score of DeepLENS is~0.404 while ORACLE reaches~0.595. Note that the mean F1 score of ORACLE is below~1 due to the existence of multiple different ground-truth summaries which cannot simultaneously match a machine-generated summary. The large gaps suggest much room for improving entity summarization techniques in future research.

\begin{figure}[!t]
	\centering
	\includegraphics[width=1\linewidth]{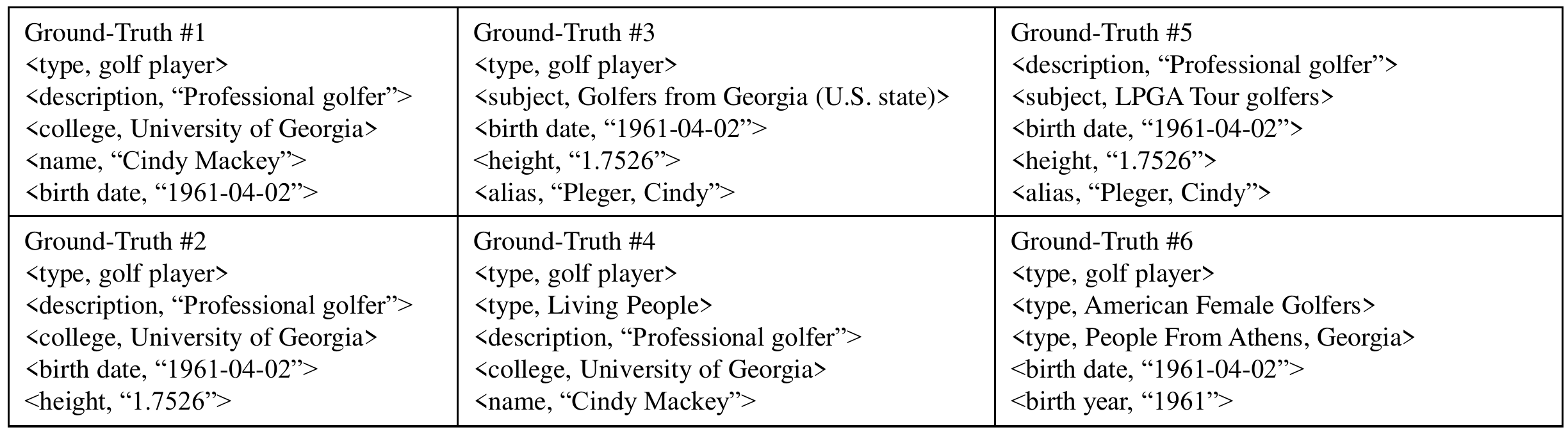}
	\caption{Ground-truth summaries under $k=5$ for entity \texttt{dbr:Cindy\_Mackey} in ESBM~v1.2.}
	\label{fig:example_gold_e52}
\end{figure}

\begin{figure}[!t]
	\centering
	\includegraphics[width=\linewidth]{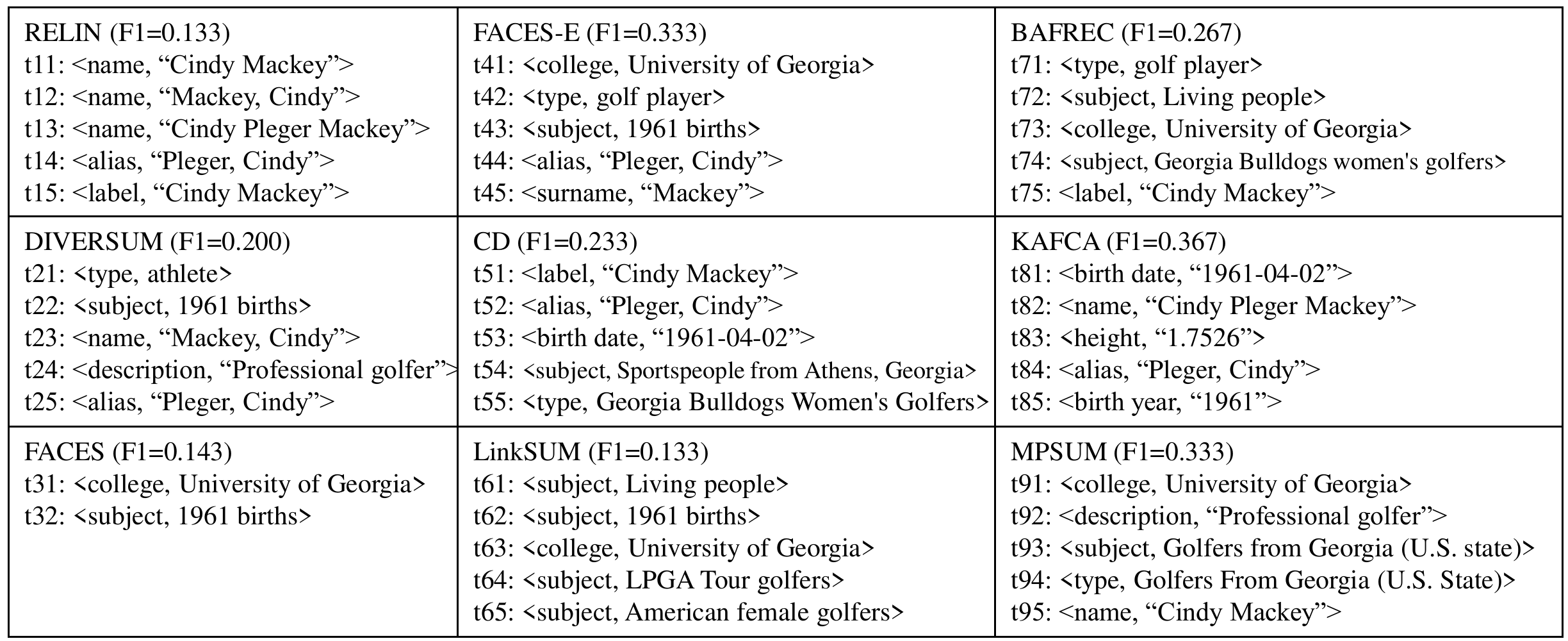}
	\caption{Summaries generated by different entity summarizers under $k=5$ for entity \texttt{dbr:Cindy\_Mackey} in ESBM~v1.2.}
	\label{fig:example_summ_e52}
\end{figure}

Moreover, we perform a case study with the entity \texttt{dbr:Cindy\_Mackey} selected from ESBM~v1.2 to illustrate and analyze the outputs of the evaluated entity summarizers. Figure~\ref{fig:example_gold_e52} shows its six human-made ground-truth summaries under $k=5$. Figure~\ref{fig:example_summ_e52} shows its nine machine-generated summaries with F1 scores. RELIN achieved the lowest F1 score. Its random surfer model selected redundant triples (t11/t12/t13/t14/t15) due to their high topical relatedness to each other. CD reduced redundancy by penalizing similar triples in its objective function and achieved a higher F1 score, but it overemphasized informativeness (t54/t55). FACES-E showed a better trade-off between informativeness and popularity (t41/t42) and obtained a higher F1 score. DIVERSUM selected diverse and locally frequent properties but paid no attention to property values and selected some less informative values (t21/t22). MPSUM selected more favorable property values (t91/t92) but still suffered redundancy (t93/t94). LinkSUM and FACES could not process type- and literal-valued triples. LinkSUM tended to select diverse and globally frequent property values but introduced many redundant properties (t61/t62/t64/t65). FACES strictly avoided selecting redundant properties from the same cluster of triples, leading to a summary comprising only two triples for this entity (t31/t32). BAFREC selected several favorable triples based on ontological informativeness (t71) and the frequency of property value (t73).
KAFCA achieved the highest F1 score. It successfully selected a triple (t81) having a locally infrequent property and a value containing a locally frequent word \texttt{"1961"}. This triple appeared in most ground-truth summaries but was rarely selected by other entity summarizers.

\begin{table}[!t]
    \caption{Extrinsic Evaluation of Entity Summarization}
	\label{table:extrinsic}
	\centering
    \resizebox{\columnwidth}{!}{
	\begin{tabular}{|l|l|l|l|}
		\hline
		& Task & Entities & Evaluation Metrics \\
		\hline
		RELIN~\cite{relin} & entity matching & DBpedia, Freebase & accuracy, time \\
		\hline
		Yovisto~\cite{yovisto} & entity search & DBpedia & success rate, questionnaire, time, recall \\
		\hline
		MMR-QSFS~\cite{sigir12} & entity search & IMDb & precision, recall \\
		\hline
		COMB~\cite{summel} & entity linking & DBpedia & accuracy, time, questionnaire \\
		\hline
	    C3D+P~\cite{c3dp} & entity matching & DBpedia, GeoNames, LinkedMDB & accuracy, time \\
		\hline
		REMES~\cite{ijcai17} & document understanding & DBpedia & questionnaire \\
		\hline
		CTab~\cite{ctab} & entity matching & DBpedia, Freebase, Wikidata, YAGO & accuracy, time, questionnaire \\
		\hline	
	\end{tabular}
	}
\end{table}

\subsection{Extrinsic Evaluation}

Table~\ref{table:extrinsic} presents known efforts in extrinsic evaluation of entity summarization. They mainly use entities in RDF data representing encyclopedic knowledge such as DBpedia, Freebase, Wikidata, and YAGO. Some use IMDb and LinkedMDB for the movie domain, or GeoNames for the geography domain. Entity summaries have been used to support a variety of downstream tasks for extrinsic evaluation. Most evaluated entity summarizers were specifically developed for the chosen task, while RELIN~\cite{relin} was designed for general purposes.

\textbf{Entity Matching.}
RELIN~\cite{relin}, C3D+P~\cite{c3dp}, and CTab~\cite{ctab} were applied to support the task of entity matching. Two or more entities were selected from different datasets as candidate matches referring to the same real-world object. Human users were invited to judge the correctness of each candidate match of entities, by comparing their summaries. An entity summarizer would be considered effective if the entity summaries it generated could help users accurately and quickly complete judgments. In~\cite{ctab}, human users were also asked to answer a post-task questionnaire about the usefulness of the presented entity summaries.

\textbf{Entity Linking.}
COMB~\cite{summel} was applied to support the task of entity linking. A set of entities (e.g.,~the Apple fruit, the Apple Inc.) were selected from DBpedia as candidates that a text span in a document (e.g.,~``apple'') might refer to. Human users were invited to select the right entity, based on the context of the text span and the summaries of the candidate entities. An entity summarizer would be considered effective if the entity summaries it generated could help users accurately and quickly complete selection. Human users were also asked to rate the usefulness of each presented summary.

\textbf{Document Understanding.}
REMES~\cite{ijcai17} was applied to support the task of document understanding. To help users understand a document such as a news article, a set of entities from DBpedia mentioned in the document were identified. Human users were invited to understand the document assisted with the summaries of these entities as background knowledge, and were asked to answer a post-task questionnaire about the usefulness of the presented entity summaries.

\textbf{Entity Search.}
Yovisto~\cite{yovisto} and MMR-QSFS~\cite{sigir12} were applied to support the task of entity search. Human users were invited to identify entities relevant to a query. Entity summaries were shown as search result snippets to support relevance judgment~\cite{sigir12}, or were provided as navigation options for exploratory search from the current entity to its related entities~\cite{yovisto}. An entity summarizer would be considered effective if the entity summaries it generated could help users accurately and quickly identify relevant entities. Accuracy was measured by information retrieval metrics such as precision and recall, and by the proportion of successfully completed tasks. In~\cite{yovisto}, human users were also asked to answer a post-task questionnaire about the usefulness of the presented entity summaries.

\section{Conclusion and Future Directions}
\label{sect:future}

We have presented a comprehensive technical review of existing entity summarizers. To identify the most salient triples of an entity, existing methods compute frequency and centrality, measure informativeness, assess diversity and coverage, and consider domains, contexts, and personalization. Multiple features are combined in different ways: from grouping and re-ranking frameworks to random surfer, combinatorial optimization, and learning to rank models. Methods using deep neural networks have also emerged. Although the research on entity summarization is progressing rapidly, the problem is still far from being solved, as suggested by the current evaluation results. Based on our review of the state of the art, we identify the following research directions for future work to overcome the limitations of existing research. Encouragingly, preliminary attempts along some directions have made promising progress.

\paragraph{Use of Semantics}

As shown in Table~\ref{table:algo}, existing entity summarizers mainly rely on statistical features such as frequency, centrality, and statistical informativeness. Indeed, they are useful when the volume of semantic data is large and rising quickly. However, it is not the magnitude but the semantics that distinguishes semantic data from other types of structured data. Entity summarizers that process semantic data are expected to exploit semantics in the data. Among others, classes and properties are associated with meaning characterized by axioms. Such ontological semantics is rich, but unfortunately, it has not been fully considered in existing research. We have only witnessed some shallow use of class hierarchies for calculating informativeness~\cite{trank,bafrec} as reviewed in Section~\ref{sect:inf} and for identifying similar triples~\cite{summel,c3dp,ensec1} as reviewed in Section~\ref{sect:div}. More semantics and more powerful reasoning capabilities can be useful for entity summarization. For example, to generate summaries for comparing two or more entities, one potentially useful method for identifying implicit inter-entity dissimilarity is to detect conflicting triples using reasoning and consistency checking techniques~\cite{DBLP:conf/esws/PaulheimS16,ggnn}.

\paragraph{Human Factors}

As shown in Table~\ref{table:algo}, most technical features in use are data-centric. They analyze various aspects of semantic data such as frequency and similarity. However, if entity summaries are generated to be presented to human users, more research attention needs to be given to human factors since the ultimate goal of entity summarization is to find triples that are truly useful for human users. For example, when summarizing the description of a book, its ISBN is statistically informative as it uniquely identifies a book. This triple is likely to be selected into an entity summary by many existing summarizers, and indeed it is useful for applications like entity resolution. However, it may not be interesting to many human users because the information it provides is not directly useful. Here, a desired feature would be to assess the meaningfulness or human friendliness of a triple to the lay audience. One potentially effective method for identifying human-friendly triples is to measure the readability of their textual forms~\cite{esster}. The measurement of text readability has been studied for decades~\cite{readability}, providing a lot of tools for reuse. Another research direction about human factors is to learn users' preference from their behavior and generate personalized entity summaries. We have seen an early attempt~\cite{ibmcdl} in Section~\ref{sect:personalization} but further studies are in demand.

\paragraph{Machine and Deep Learning}

As reviewed in Section~\ref{sect:ltr}, only a few entity summarizers use supervised methods~\cite{trank,sigir12,dynes,eventkg}. These methods focus on specific applications where labeled data is available for training. For generic entity summarization, it is expensive to manually label a large set of entity summaries as training data. This has hindered the development of supervised entity summarization in a general setting. Recent benchmarks for intrinsic evaluation of entity summarization have made available some ground-truth entity summaries. Based on that, new supervised entity summarizers have emerged, including a few deep learning based methods~\cite{esa,deeplens} as reviewed in Section~\ref{sect:deep}. However, compared with the fruitful research on supervised document summarization~\cite{ts1,ts2}, there is still plenty of room for applying machine and deep learning techniques to entity summarization. To obtain more labeled entity summaries for training, one possible solution is to programmatically generate labels. For example, in~\cite{ipm}, entity descriptions in DBpedia are automatically aligned with abstracts of DBpedia articles to identify key triples for each entity and form a summary. A large number of labeled entity summaries can be generated in this way, though containing noise and hence the supervision would be weak~\cite{DBLP:conf/ijcai/Li0LZKGC20}.

\paragraph{Non-Extractive Methods}

All the entity summarizers reviewed in this article use extractive methods. They generate a summary for an entity description by extracting a subset of triples. In document summarization~\cite{ts1,ts2}, a paradigm complementary to extractive methods is non-extractive summarization, which generates an abstractive summary consisting of ideas or concepts that are taken from the original description but are re-interpreted in different and better forms. For document summarization, non-extractive methods are complex as they need extensive natural language processing. For entity summarization, the exploration of non-extractive methods is in the initial phase. It is still open to discuss the concrete form of an abstractive entity summary. One promising direction is to generate a textual summary, as text may be more readable than triples. A textual summary can either be generated from a computed triple-structured summary using natural language generation techniques~\cite{DBLP:journals/jair/GattK18}, or be directly generated from an entity description using sequence-to-sequence models~\cite{ed2t1,ed2t2,ed2t3} which combine triple selection and text generation into a hybrid process.

\paragraph{Interactive Methods}

Entity summarizers are not perfect at all times. Indeed, as reviewed in Section~\ref{sect:intrinsic-results}, they are still far from perfect. When an one-shot summary fails to fulfill a user's information needs, the user may expect to see an improved summary after interacting with the summarizer. This kind of interactive entity summarization has not received much research attention. For document summarization, interactive methods solicit feedback from the user to capture opinions and interests~\cite{its1,its2}. Entity summarization can follow this line of research, and user feedback can be positive or negative opinions about triples. However, considering the structured and semantic nature of entity descriptions, directly adapting text-based methods for interactive document summarization is unlikely to be effective. Research is needed for finding suitable models for characterizing a user's interaction with an entity summarizer. In a pioneer work~\cite{feedbackeswc}, reinforcement learning is employed to model interactive entity summarization where a user can provide negative feedback on presented triples and then be served with an improved summary. Apart from that, more effective methods may be developed by incorporating state-of-the-art interactive information retrieval approaches such as online learning to rank~\cite{DBLP:conf/sigir/JagermanOR19}.
\section{Acknowledgements}
This work was supported by the NSFC [grant number 62072224 and 61772264].





\bibliographystyle{elsarticle-num}
\bibliography{jwsbibfile}







\end{document}